\documentclass[aip,reprint,amsmath,amssymb,unsortedaddress,superscriptaddress]{revtex4-1}
\usepackage{color}
\usepackage{graphicx}
\usepackage[]{enumerate}
\usepackage{bm}
\usepackage{subfig}
\usepackage{amsmath}

\DeclareBoldMathCommand{\bV}{V}
\DeclareBoldMathCommand{\bv}{v}
\DeclareBoldMathCommand{\bu}{u}
\DeclareBoldMathCommand{\bx}{x}
\DeclareBoldMathCommand{\by}{y}
\DeclareBoldMathCommand{\bz}{z}
\DeclareBoldMathCommand{\br}{r}
\DeclareBoldMathCommand{\bb}{b}
\DeclareBoldMathCommand{\be}{e}
\DeclareBoldMathCommand{\bB}{B}
\DeclareBoldMathCommand{\bE}{E}
\DeclareBoldMathCommand{\bk}{k}
\DeclareBoldMathCommand{\bA}{A}
\DeclareBoldMathCommand{\bJ}{J}
\DeclareBoldMathCommand{\bR}{R}

\begin{document}
\title{Interpreting Magnetic Variance Anisotropy Measurements in the Solar Wind}
\author{J. M. TenBarge}
\email{jason-tenbarge@uiowa.edu}
\affiliation{Department of Physics and Astronomy, University of Iowa, Iowa City, IA, USA}
\author{J. J. Podesta}
\affiliation{Space Science Institute, Boulder, CO, USA}
\author{K. G. Klein and G. G. Howes}
\affiliation{Department of Physics and Astronomy, University of Iowa, Iowa City, IA, USA}

\begin{abstract}
The magnetic variance anisotropy ($\mathcal{A}_m$) of the solar wind has been used widely as a method to identify the nature of solar wind turbulent fluctuations; however, a thorough discussion of the meaning and interpretation of the $\mathcal{A}_m$ has not appeared in the literature. This paper explores the implications and limitations of using the $\mathcal{A}_m$ as a method for constraining the solar wind fluctuation mode composition and presents a more informative method for interpreting spacecraft data. The paper also compares predictions of the $\mathcal{A}_m$ from linear theory to nonlinear turbulence simulations and solar wind measurements. In both cases, linear theory compares well and suggests the solar wind for the interval studied is dominantly Alfv\'{e}nic in the inertial and dissipation ranges to scales $k \rho_i \simeq 5$.
\end{abstract}


\maketitle
\section{Introduction}\label{sec:intro}

\textit{In situ} measurements of solar wind turbulence consistently show one dimensional magnetic energy spectra that obey a broken power law, typically having spectral indices $-5/3$ in the inertial range and steepening in the dissipation range \citep{Tu:1995,Bruno:2005,Alexandrova:2011}. The inertial and dissipation ranges correspond to scales $k r_i < 1$ and $k r_i > 1$ respectively, where $k$ is the wavenumber and $r_i$ is the relevant ion kinetic scale, typically either the ion gyroradius or inertial length. 

Although measurements of the magnetic energy spectrum are common, the spectrum alone does not provide direct insight into the nature of solar wind fluctuations. Since the solar wind turbulence is electromagnetic, the expectation is that the solar wind fluctuations will exhibit characteristics of the three basic electromagnetic plasma wave modes at the large scales of the inertial range: Alfv\'{e}n, fast magnetosonic, and slow magnetosonic \citep{Klein:2012}. Similarly, the dissipation range is expected to be populated by the kinetic scale counterparts of the three wave modes.

Based upon a variety of metrics, the Alfv\'{e}n mode appears to be the dominant wave mode in the inertial range \citep{Tu:1995,Bruno:2005,Horbury:2008,Podesta:2009a,Howes:2011a}. The composition of solar wind fluctuations in the dissipation range is less well constrained due to the dearth of high frequency measurements in the free solar wind necessary to probe this region, but recent observations suggest the kinetic Alfv\'{e}n wave (KAW) is the dominant mode in the free solar wind \citep{Bale:2005,Chandran:2009b,Sahraoui:2010b,Podesta:2011a,Salem:2012}---the KAW is the dissipation range extension of the inertial range Alfv\'{e}n wave in the $k_\perp > k_\parallel$ region of wavenumber space, where parallel and perpendicular are with respect to the local mean magnetic field, $B_0$ .

One of the commonly used metrics is the magnetic variance anisotropy ($\mathcal{A}_m$), first introduced by \citet{Belcher:1971}. The $\mathcal{A}_m$ is a measure of the fluctuation anisotropy and has come to be defined as $\mathcal{A}_m = \langle|\delta B_\perp|^2\rangle/\langle|\delta B_\parallel|^2\rangle$, where angle brackets indicate averages, the $\delta B$ are fluctuating quantities about the local mean magnetic field, and $\delta B_\perp^2$ is the total energy in the plane perpendicular to the local mean magnetic field. It is important to not confuse this quantity with the wavevector anisotropy inherent to and often discussed in plasma turbulence: the wavevector anisotropy and the $\mathcal{A}_m$ are not directly related quantities \citep{Matthaeus:1995}. Physically, the $\mathcal{A}_m$ can be viewed as a measure of the magnetic compressibility of the plasma, $C_\parallel = \langle|\delta B_\parallel|^2\rangle / \langle|\delta \bB|^2\rangle$, since $\mathcal{A}_m + 1= 1 / C_\parallel$.

The version of the $\mathcal{A}_m$ first introduced by \citet{Belcher:1971} has been expanded upon in recent papers using the $\mathcal{A}_m$ defined above, e.g., \citep{Leamon:1998a,Smith:2006a,Hamilton:2008,Gary:2009,He:2012,Podesta:2012,Salem:2012,Smith:2012}.  We attempt here to provide a discussion that includes a theoretical basis for the interpretation of $\mathcal{A}_m$ measurements in the solar wind. In \S \ref{sec:waves}, we describe in detail the expected behaviour of the $\mathcal{A}_m$ of the three constituent electromagnetic wave modes in the solar wind inertial range and discuss their transition into the dissipation range. In \S \ref{sec:combos}, we discuss the effect of superposing the three wave modes. \S \ref{sec:measure} explores alternative procedures for constructing the $\mathcal{A}_m$, and compares the linear theory prediction from \S \ref{sec:waves} to fully nonlinear turbulence simulations as a means establish the validity of linear theory to nonlinear turbulence. \S \ref{sec:solar} reviews some of the recent uses of the $\mathcal{A}_m$ to quantify the composition of solar wind fluctuations and presents new measurements of the $\mathcal{A}_m$ from the Stereo A spacecraft. 


\section{Wave Modes}\label{sec:waves}
\begin{figure}[t]
		\includegraphics[width=\linewidth]{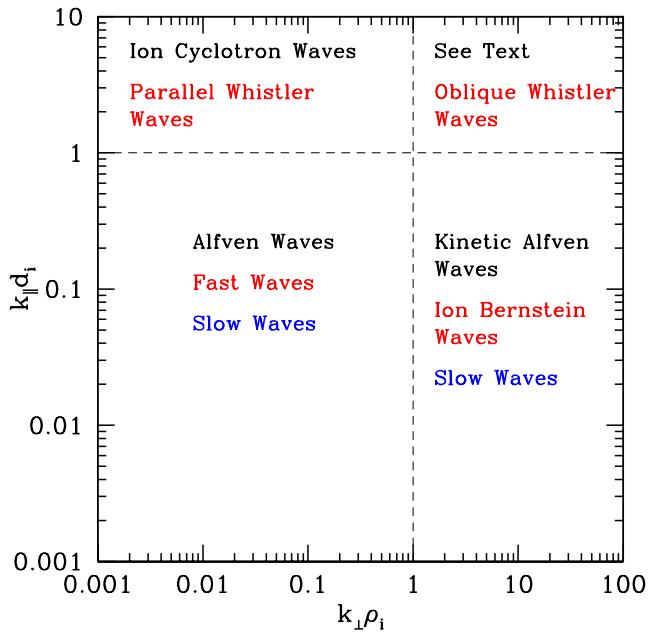}
      \caption{(Color online) Schematic diagram of wave mode transitions in wavenumber space. Black, red, and blue indicate Alfv\'{e}n, fast, and slow mode branches respectively.}\label{fig:waves}
\end{figure}

We begin by enumerating the properties of the three linear wave modes which are the collisionless counterparts to the Alfv\'{e}n and fast and slow magnetosonic modes in compressible MHD \citep{Klein:2012}. Since the solar wind is a weakly collisional plasma, we focus here on the roots provided by the collisionless Vlasov-Maxwell (VM) system of equations, which are a function of $k_\parallel, k_\perp$, $\beta_i$, $T_i / T_e,$ and $v_{ti} / c$, where $v_{ti}^2 = 2 T_i / m_i$ is the ion (protons only) thermal speed. $v_{ti} /   c = 10^{-4}$ and $T_i = T_e$ unless otherwise stated. Figure \ref{fig:waves} presents a schematic diagram summarizing the nomenclature of the three wave branches in different regions of wavenumber space.

\subsection{Alfv\'{e}n Mode}
At scales $k_\perp \rho_i \ll 1$ and $k_\parallel d_i \ll 1$ in the free solar wind, the Alfv\'{e}n wave is the dominantly observed wave mode \citep{Tu:1995,Bruno:2005,Horbury:2008,Podesta:2009a,Sahraoui:2010b,Podesta:2011a}, where $\rho_i = v_{ti} / \Omega_i$ is the ion (proton) gyroradius, $\Omega_i = e B_0 / m_i c$ is the proton gyrofrequency, $d_i = c / \omega_{pi} = \rho_i / \sqrt{\beta_i}$ is the ion inertial length, and $\omega_{pi}^2 = 4 \pi n_i e^2 / m_i$ is the ion plasma frequency. The turbulent energy cascade at these scales has been thoroughly explored in the literature, where the one dimensional perpendicular magnetic energy is observed to scale as $E_{B_\perp} \propto k_\perp^{-\alpha}$ with a wavevector anisotropy $k_\parallel \propto k_\perp^\xi$. When the turbulence is in critical balance---i.e, the nonlinear cascade rate is of order the linear  frequency---the theoretical values for $\alpha$ and $\xi$ are expected to be $\alpha = 5/3$ and $\xi = 2/3$ for the model of Goldreich-Sridhar \citep{Goldreich:1995} or $\alpha = 3/2$ and $\xi = 1/2$ for the dynamic alignment model \citep{Boldyrev:2005,Boldyrev:2006}. Which model is correct is not completely settled. Solar wind observations typically show a magnetic field spectrum with $\alpha \simeq 5/3$ and total (kinetic plus magnetic) energy spectrum with $\alpha \simeq 3/2$ \citep{Luo:2010,Podesta:2010b,Wicks:2010a,Boldyrev:2011,Chen:2011a,Chen:2011}, while MHD turbulence simulations suggest the dynamic alignment model may be more correct \citep{Mason:2006,Mason:2008}. For the purpose of modelling the energy cascade, we assume the Goldreich-Sridhar model for simplicity. The choice of model does not significantly affect the behaviour of the $\mathcal{A}_m$.

\begin{figure*}[t]
\begin{center}	
	\subfloat[]{\label{fig:betas}\includegraphics[width=0.5\textwidth]{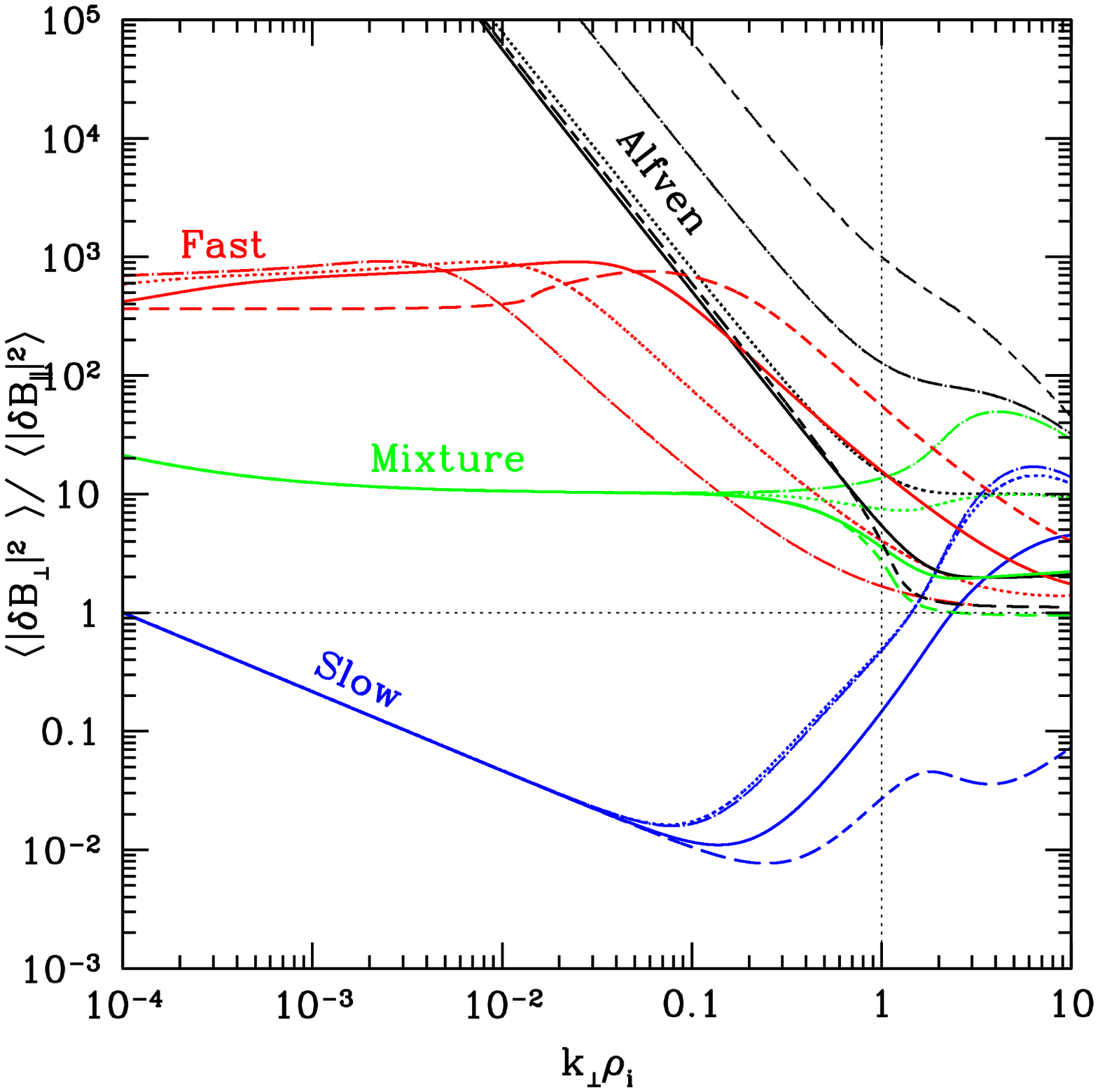}}                
 	\subfloat[]{\label{fig:betas_freq}\includegraphics[width=0.5\textwidth]{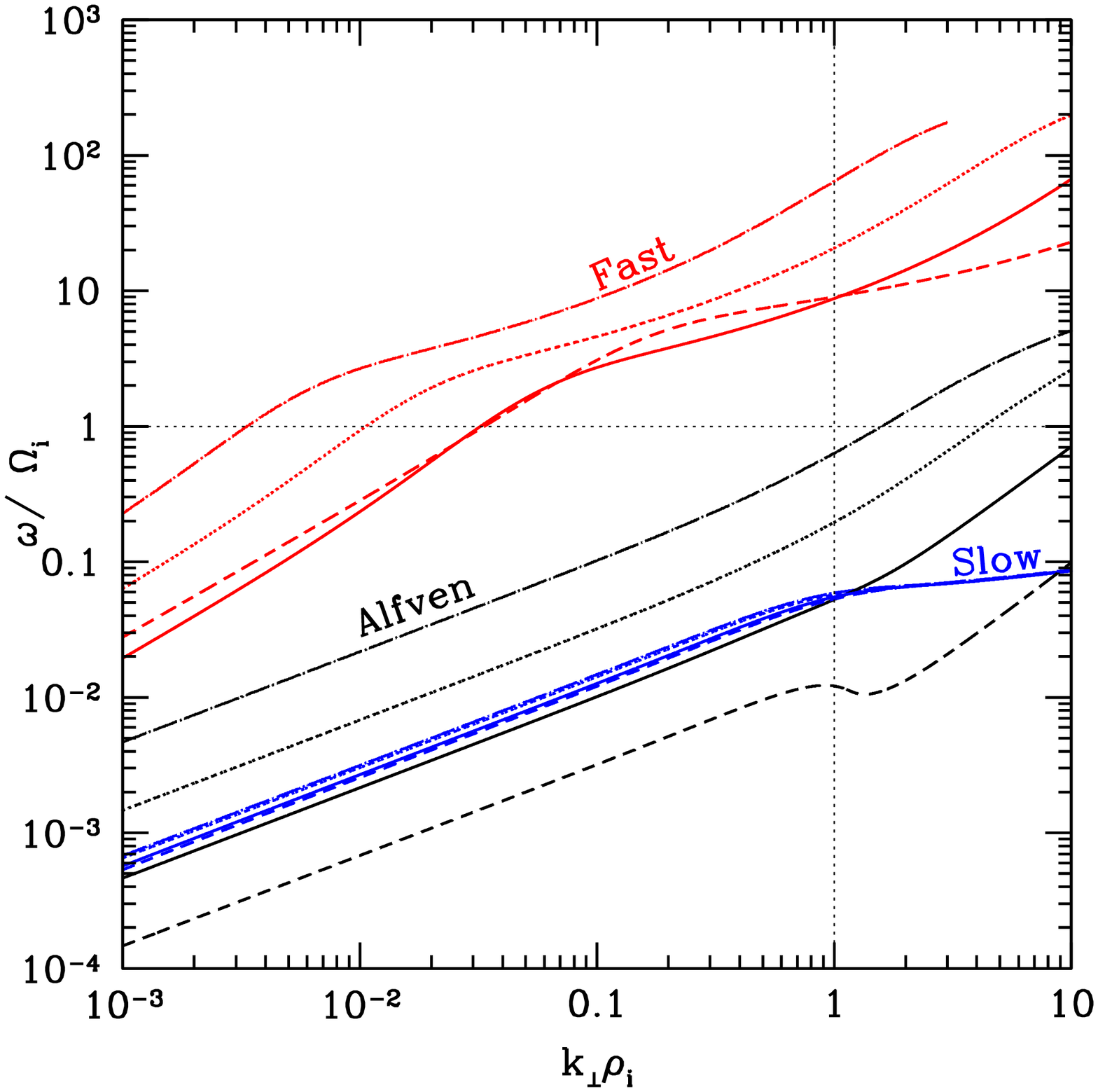}}
\end{center}
  \caption{(Color online) (a) The predicted variance anisotropy based on linear Vlasov-Maxwell theory  for the Alfv\'{e}n (black), fast (red), and slow (blue) roots as a function of wavenumber. Green lines represent a fiducial solar wind mixture of the three modes composed of $90\%$ Alfv\'{e}n, $9\%$ slow, and $1\%$ fast waves. Dot-dashed, dotted, solid, and dashed lines represent $\beta_i = 0.01$, $0.1$, $1$, and $10$ solutions respectively. The Alfv\'{e}n and slow roots are assumed to have a spectral anisotropy given by critical balance with an outer scale $k_i \rho_i = 10^{-4}$ and the fast root has $\theta_{kB} = 3^\circ$ up to $k_\parallel d_i = 1$ and $k_\parallel \propto k_\perp^{1/3}$ thereafter. (b) The dispersion relations of the same wave modes.}
\end{figure*}

The properties of the MHD Alfv\'{e}n root are well understood. However, the MHD solution of the Alfv\'{e}n root is incompressible since $\delta B_\parallel = 0$, suggesting the Alfv\'{e}nic $\mathcal{A}_m$ is unbounded. The full collisionless VM solution of the Alfv\'{e}n root in the inertial range has a small but non-vanishing $\delta B_\parallel$ which increases with $k_\perp$, thereby causing the $\mathcal{A}_m$ to decrease with $k_\perp$. The VM solution for the Alfv\'{e}n root (black) $\mathcal{A}_m$ and dispersion relation are plotted against $k_\perp \rho_i$ in Figures \ref{fig:betas} and \ref{fig:betas_freq} for $\beta_i = 0.01, 0.1, 1$, and $10$ (dash-dotted, dotted, solid, and dashed respectively), $\beta_i = v_{ti}^2 / v_A^2$ and $v_A^2 = B_0^2 / 4 \pi m_i n$ is the Alfv\'{e}n speed. The solution assumes an inertial range spectral anisotropy as described by the Goldreich-Sridhar model, $k_\parallel = k_i^{1/3} k_\perp^{2/3}$, where $k_i$ is the isotropic outer-scale of the turbulence and is taken to be $k_i \rho_i = 10^{-4}$, consistent with \textit{in situ} solar wind measurements at $1$~AU \citep{Howes:2008b}.

Continuing along the critical balance cascade to kinetic scales naturally produces spectral anisotropy with $k_\perp \gg k_\parallel$. So, we next consider the Alfv\'{e}n root with $k_\perp \rho_i > 1$ and $k_\parallel d_i < 1$. At these scales, the Alfv\'{e}n wave transitions into the kinetic Alfv\'{e}n wave (KAW). The KAW is dispersive, damped, and much more compressible than the Alfv\'{e}n wave. The dispersive nature of the root steepens the magnetic energy spectrum to $\alpha = 7/3$ in the undamped case \citep{Howes:2008b,Schekochihin:2009}, while the inclusion of damping steepens the spectrum further \citep{Howes:2011c,Howes:2011b,TenBarge:2012c}. Damping here refers to collisionless wave-particle interactions, primarily ion transit time damping on the parallel magnetic field that peaks at ion scales for $\beta_i \gtrsim 1$ \citep{Barnes:1966,Quataert:1998} and electron Landau damping on the parallel electric field that peaks at electron scales \citep{Howes:2006,Schekochihin:2009}. The spectral anisotropy of the KAW is assumed to scale as $k_\parallel \propto k_\perp^{1/3}$ \citep{Galtier:2006,Howes:2008b,Schekochihin:2009,TenBarge:2011a}. 

\begin{figure}[t]
		\includegraphics[width=\linewidth]{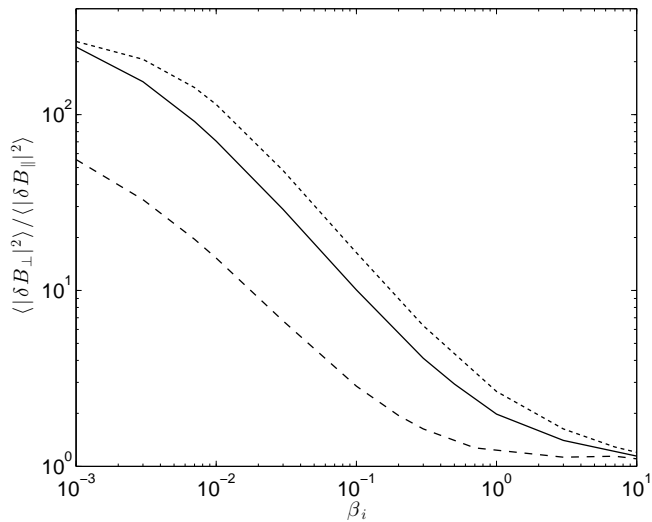}
      \caption{The $\beta_i$ dependence of the KAW root for $T_i / T_e = 10$, $1$, and $0.1$ (dotted, solid, and dashed respectively) averaged across $k_\perp \rho_i \in [3,4]$.}\label{fig:kaw_beta_dep}
\end{figure}

The increased compressibility of the KAW can be seen at scales $k_\perp \rho_i > 1$ in Figure \ref{fig:betas}, where the KAW $\mathcal{A}_m$ is seen to have a $\beta_i$ dependent plateau. The KAW $\mathcal{A}_m$ in the dissipation range also depends upon the ion-to-electron temperature ratio, $T_i / T_e$. An analytical form for the KAW $\mathcal{A}_m$ can be derived (see Appendix \ref{app:ermhd}) within the framework of electron reduced MHD (ERMHD) developed in \citet{Schekochihin:2009},
\begin{equation}
\mathcal{A}_m = \frac{2 + \beta_i \left(1+T_e/T_i\right)}{\beta_i \left(1+T_e/T_i\right)}.
\end{equation}
The KAW $\mathcal{A}_m$ has no wavenumber dependence and thus plateaus at a value determined by $\beta_i$ and $T_i / T_e$. The $\beta_i$ and temperature ratio dependence of the KAW $\mathcal{A}_m$ derived from VM theory is plotted in Figure \ref{fig:kaw_beta_dep}, where the value for the $\mathcal{A}_m$ is averaged across the plateau at $k_\perp \rho_i \in [3,4]$.

The behaviour of the Alfv\'{e}n root becomes more complicated for scales $k_\parallel d_i > 1$. When $k_\parallel d_i > 1$ and $k_\perp \rho_i \lesssim 1$, the Alfv\'{e}n root becomes the Alfv\'{e}n ion cyclotron mode. This mode is strongly cyclotron damped and characterized by a left-handed magnetic helicity and left-handed electric field polarization. Because of the strong damping and \textit{in situ} solar wind observations suggesting this mode is less common than the KAW \citep{Podesta:2011a,He:2012}, we do not consider it further. When $k_\parallel d_i \gtrsim 1$ and $k_\perp \rho_i \gtrsim 1$, the behaviour of the Alfv\'{e}n root has not been thoroughly explored. The ion Bernstein wave (IBW) \citep{Stix:1992} is conventionally assumed to couple to the KAW at the ion gyrofrequency; however, the KAW root that exists for $k_\parallel d_i < 1$ may continue undamped at higher frequencies \citep{Howes:2008b,Sahraoui:2011,Klein:2012a}. Since the behaviour in this region of wavenumber space is uncertain, we do not attempt to describe it here. 

\subsection{Slow Mode}

The majority of \textit{in situ} solar wind observations of the inertial range suggest the component responsible for most of the measured free solar wind compressibility are pressure balanced structures (PBSs) \citep{Tu:1995,Bruno:2005}; however, linear PBSs are degenerate with the $k_\parallel = 0$, non-propagating limit of slow waves \citep{Tu:1995,Kellogg:2005}. For simplicity, we classify linear PBSs as slow modes. The association of the compressible portion of the solar wind to PBSs is due to the measured anti-correlation of the thermal and magnetic pressure \citep{Burlaga:1970} or anti-correlation of density and magnetic field magnitude \citep{Vellante:1987,Roberts:1990} at inertial range scales. Recent analyses \citep{Howes:2011a,Klein:2012} exploring the more telling anti-correlation of density and parallel magnetic field suggest that the compressible portion of the solar wind is primarily composed of propagating slow modes rather than PBSs. These analyses suggest that on average Alfv\'{e}n modes comprise $\sim 90\%$ of the energy and slow modes $\sim 10\%$ of the energy.

The notion that the warm, $T_i /T_e \simeq 1$, $\beta_i \simeq 1$ propagating slow mode exists in the solar wind defies conventional wisdom that the slow mode is strongly damped in such plasmas \citep{Barnes:1966}, so we here elucidate this point. Theory \citep{Goldreich:1997,Lithwick:2001,Schekochihin:2009} and numerical simulations \citep{Cho:2002a,Cho:2003} suggest the slow mode does not have its own active turbulent cascade; rather, the slow mode is passively cascaded by the Alfv\'{e}nic turbulence. The slow mode damping rate is proportional to the parallel wavenumber, $\gamma_S \propto k_{\parallel S}$, and the strong damping of the slow mode suggests $\gamma_S / \omega_S \simeq 1$, where $k_{\parallel S}$, $\gamma_S$, and $\omega_S$ are the parallel wavenumber, linear damping rate, and frequency of the slow mode. However, the passive cascade of the slow mode implies that the slow modes are cascaded by the Alfv\'{e}n cascade on the Alfv\'{e}n timescale, $\omega_A = k_{\parallel A} v_A$. Therefore, at a given $k_\perp^*$, the parallel wavenumber of the slow mode compared to the parallel wavenumber of the Alfv\'{e}n mode  determines the strength of the damping relative to the cascade rate: at a particular scale, $k_\perp^*$, a slow mode with $k_{\parallel S}^* \ll k_{\parallel A}^*$ can be passively cascade before being damped since the slow mode damping rate will be smaller than the Alfv\'{e}n cascade time, $\gamma^*_S / \omega_A^* \ll 1$.



For slow waves in the MHD limit, $\mathcal{A}_m = k_\parallel^2 / k_\perp^2$ (see Appendix \ref{app:mhd} for the derivation of this equation), and the VM solution does not deviate significantly from the MHD solution until finite Larmor radius effects become dominant at $k_\perp \rho_i \simeq 0.1$. The VM solution for the slow mode $\mathcal{A}_m$ (blue) is plotted in Figure \ref{fig:betas} assuming a passive cascade of slow waves with $k_{\parallel S} \propto k_{\perp S}^{2/3}$---a more anisotropic cascade decreases the $\mathcal{A}_m$ below that in the Figure but does not alter the qualitative behaviour. The $\mathcal{A}_m$ for a PBS, i.e., $k_\parallel = 0$ slow mode, is identically zero since $\delta B_\perp = 0$ for these modes. 

Since the slow mode is strongly damped at large $k_\parallel$ unless $T_i \ll T_e$, we will exclude a discussion of the behaviour of this root for $k_\parallel d_i > 1$.

\subsection{Fast Mode}\label{sec:fastmode}


At inertial range scales, the fast mode is well described by MHD, whose solution provides an $\mathcal{A}_m$ identical to that of slow waves: $\mathcal{A}_m = k_\parallel^2 / k_\perp^2$. However, the distribution of fast waves in wavenumber space is not as well constrained as that of slow modes since the fast mode is not strongly damped for parallel propagation. Further, compressible MHD turbulence simulations indicate that the fast mode is cascaded isotropically in wavenumber space \citep{Cho:2003}. Therefore, a turbulent cascade of fast modes has a $\mathcal{A}_m$ that can take on all possible values, and the measured average value of the $\mathcal{A}_m$ will depend sensitively on the distribution of fast modes in wavenumber space. To represent the fast mode turbulence, we plot in Figure \ref{fig:betas} the $\mathcal{A}_m$ for a fast wave (red) with $\theta_{kB} = \arccos{\left(\bk \cdot \bB_0 / |\bk| |\bB_0|\right)} = 3^\circ$. We choose this value as representative because the average $\mathcal{A}_m$ of an isotropic distribution of fast modes will be dominated by those modes with $\theta_{kB} \simeq 0$ and the largest measured $\mathcal{A}_m$ in a large ensemble of solar wind data is $\sim 500$ \citep{Smith:2006a}.

In the dissipation range at scales $k_\parallel d_i > 1$, the fast mode transitions to a parallel whistler ($k_\perp \rho_i < 1$) or an oblique whistler ($k_\perp \rho_i > 1$) wave with $\omega / \Omega_i > 1$. The whistler mode is well described by the electron MHD (EMHD) equations \citep{Kingsep:1990}, which describes phenomena on scales $k d_i > 1$. Note, one must take care when applying the EMHD equations, because EMHD is only valid for $T_i  \ll T_e$. For scales $k_\perp d_i > 1$ and $k_\parallel d_i < 1$, EMHD describes the cold ion limit of KAWs and not whistler waves \citep{Ito:2004,Hirose:2004,Howes:2009b,Schekochihin:2009}. Solving the EMHD equations provides $\mathcal{A}_m = 1 + 2 k_\parallel^2 / k_\perp^2$ (see Appendix \ref{app:emhd}), which is a good approximation of the VM solution and can again attain all values between $1$ and infinity and is sensitively dependent upon the wavenumber distribution. Therefore, if the propagation angle or whistler wavenumber distribution does not change from that of the inertial range fast modes, the average $\mathcal{A}_m$ will approximately double in the dissipation range. However, theory \citep{Galtier:2003,Cho:2004,Narita:2010a} and simulation \citep{Dastgeer:2000,Cho:2004,Svidzinski:2009,Saito:2010} suggest the cascade of whistler waves is highly anisotropic in the same sense as critically balanced KAWs, $k_\parallel \propto k_\perp^{1/3}$. The anisotropic nature of the dissipation range cascade suggests the average $\mathcal{A}_m$ of a distribution of whistler waves will decrease rapidly to a $\beta_i$ independent value $\mathcal{A}_m \sim 1$ as the cascade progresses to smaller scales. The whistler portion of the fast modes plotted in Figure \ref{fig:betas} follows the critical balance prediction described above, with $\theta_{kB} = 3^\circ$ up to $k_\parallel d_i = 1$ and then $k_\parallel \propto k_\perp^{1/3}$. Note that the $\mathcal{A}_m$ of whistler waves has a very weak $\beta_i$ and ion-to-electron temperature ratio dependence. The apparent $\beta_i$ dependence of the whistler branch in Figure \ref{fig:betas} is due to the fast-whistler break point being at $k_\parallel d_i = k_\parallel \rho_i / \sqrt{\beta_i}$; therefore, the $\beta_i$ dependence in the Figure would vanish if the x-axis were normalized to the ion inertial length.

\begin{figure*}[t]
\begin{center}	
	\subfloat[]{\label{fig:beta1_whistler_var}\includegraphics[width=0.5\textwidth]{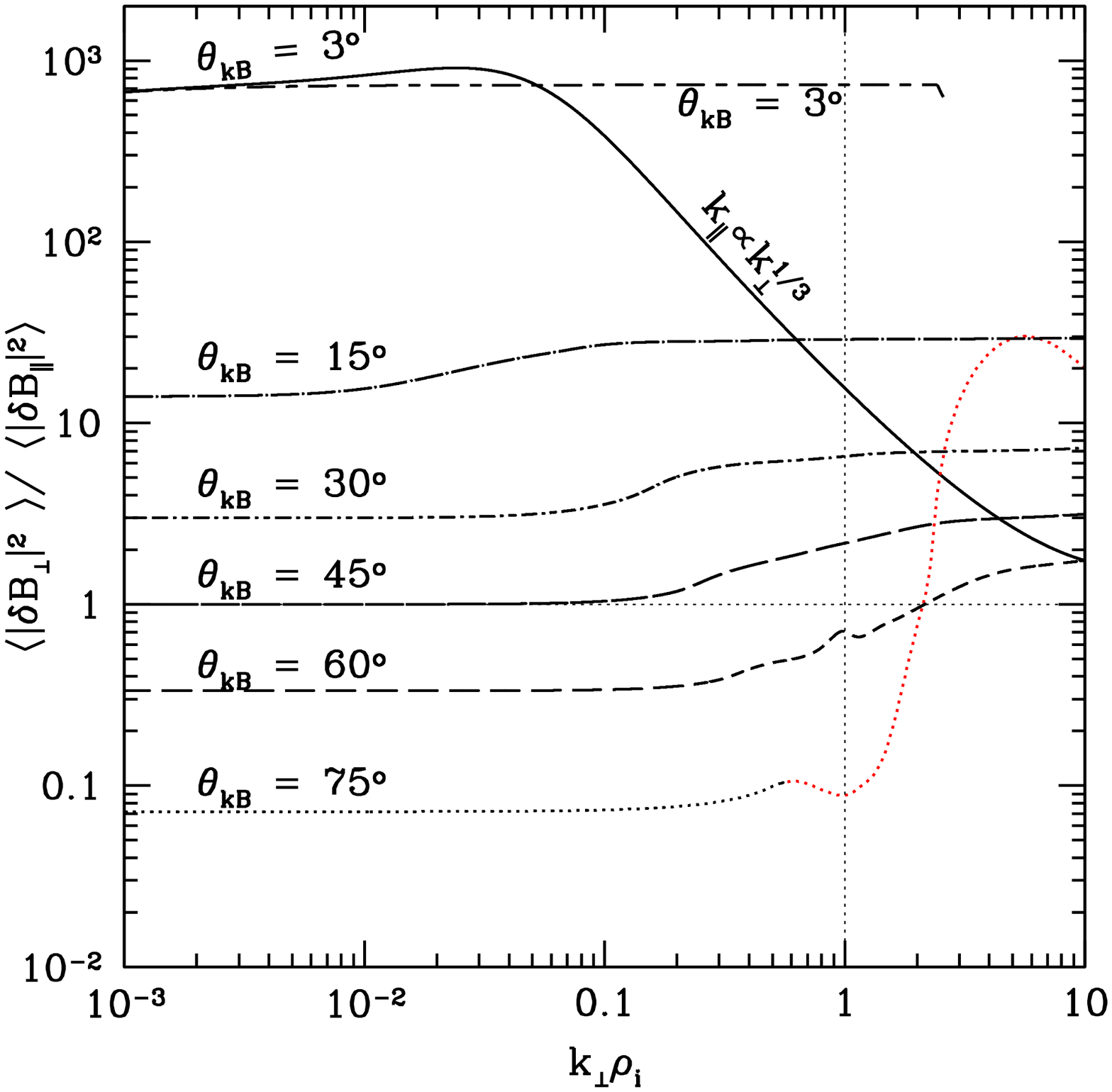}}                
 	\subfloat[]{\label{fig:beta1_whistler_freq}\includegraphics[width=0.5\textwidth]{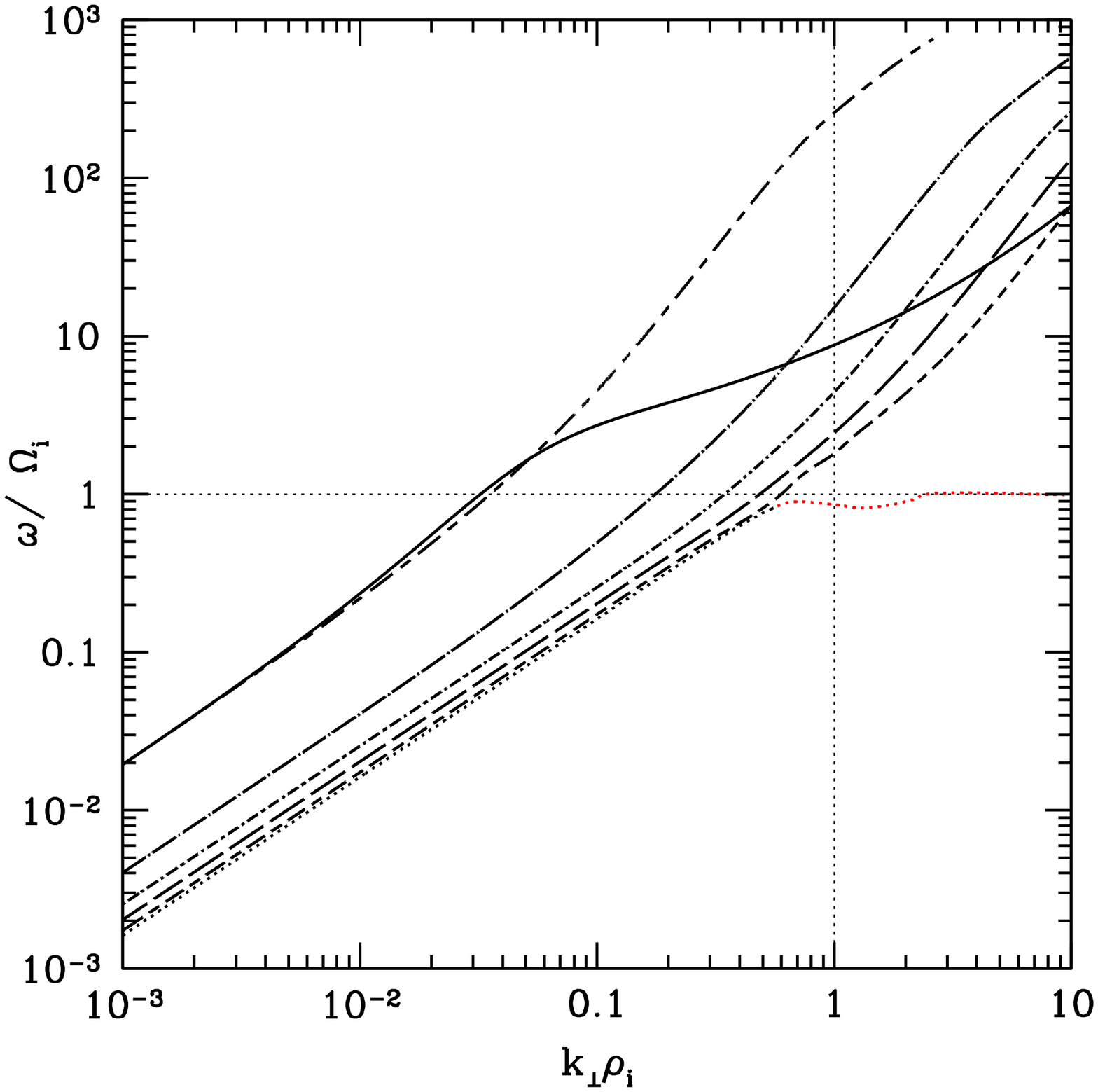}}
\end{center}
  \caption{(Color online) (a) The predicted variance anisotropy based on linear Vlasov-Maxwell theory with $\beta_i = 1$ and $T_i = T_e$ for the fast/whistler root for different propagation angles, $\theta_{kB} = 3^\circ$, $15^\circ$, $30^\circ$, $45^\circ$, $60^\circ$, $75^\circ$ (long-short dashed, long dash-dotted, short dash-dotted, long dashed, short dashed, dotted), and $3^\circ$ transitioning to critically balanced whistler (solid). When the $\theta_{kB} = 75^\circ$ root transitions to an IBW, the color changes to red. The $\theta_{kB} = 3^\circ$ root terminates at $k_\perp \rho_i \simeq 3$ due to the solver's inability to resolve the high frequency of the root. (b) The dispersion relation for each root.}
\end{figure*}

For completeness, the fast mode to whistler transition $\mathcal{A}_m$ is shown for several different propagation angles, $\theta_{kB}$, in Figure \ref{fig:beta1_whistler_var}. The $\mathcal{A}_m$ agrees well with the predictions above, except in the $\theta_{kB} = 75^\circ$ case. For such highly oblique angles, the fast mode transitions to an IBW: at scales $k_\perp \rho_i \gtrsim 1$ and $k_\parallel d_i \lesssim 1$, the fast mode transitions to the electrostatic IBW with frequency approximately equal to an integer multiple of the ion cyclotron frequency \citep{Stix:1992,LiHabbal:2001,Howes:2009b}. The transition to the IBW is clear in the fast/whistler wave dispersion relations in Figure \ref{fig:beta1_whistler_freq} corresponding for the same angles as the $\mathcal{A}_m$ presented in \ref{fig:beta1_whistler_var}. Although the IBW is electromagnetic during the transition from electromagnetic fast or Alfv\'{e}n modes, IBWs are dominantly electrostatic \citep{Stix:1992}. The properties of the transition electromagnetic IBW have not been fully explored in the literature, so we will not consider the potential contribution of IBWs to the dissipation range $\mathcal{A}_m$.
 

The wavenumber distribution of whistler waves generated in a local instability such as the parallel firehose \citep{Gary:1998} and whistler anisotropy instabilities is not easily determined and depends upon the instability, so we will not consider them further.

\section{Combinations of Modes and $\mathcal{A}_m$}\label{sec:combos}

Assuming the solar wind consists of combinations of the three wave modes described in the preceding section both in the inertial and dissipation ranges, we need to consider the effect of superposing the modes. The three linear modes can be combined into a total $\mathcal{A}_m$ measure 

\begin{equation}\label{eq:MVA_full}
\begin{split}
&\mathcal{A}_{mT} = \\ 
 &\frac{AC~C_{\perp A} + (1-AC) \left[FS~C_{\perp F} + (1-FS) C_{\perp S}\right]}{AC~C_{\parallel A} + (1-AC) \left[FS~C_{\parallel F} + (1-FS) C_{\parallel S}\right]},
\end{split}
\end{equation} 
where $AC$ is the fraction of Alfv\'{e}n to total energy, $FS$ is the fraction of fast to total compressible (fast plus slow) energy, $C_\perp = \langle|\delta B_\perp|^2\rangle / \langle|\delta \bB|^2\rangle$, and subscripts A, F, and S refer to Alfv\'{e}n, fast, and slow modes. Note that $AC \in [0,1]$ and $FS \in [0,1]$. $C_\parallel$ and $C_\perp$ for each mode can both be written in terms of the $\mathcal{A}_m$ as
\begin{equation}\label{eq:para_comp}
C_\parallel = \frac{1}{1 + \mathcal{A}_m}
\end{equation}
and
\begin{equation}
C_\perp = \frac{\mathcal{A}_m}{1 + \mathcal{A}_m}.
\end{equation}

\subsection{Inertial Range}\label{sec:MVA_inertial}
\begin{figure}[t]
		\includegraphics[width=\linewidth]{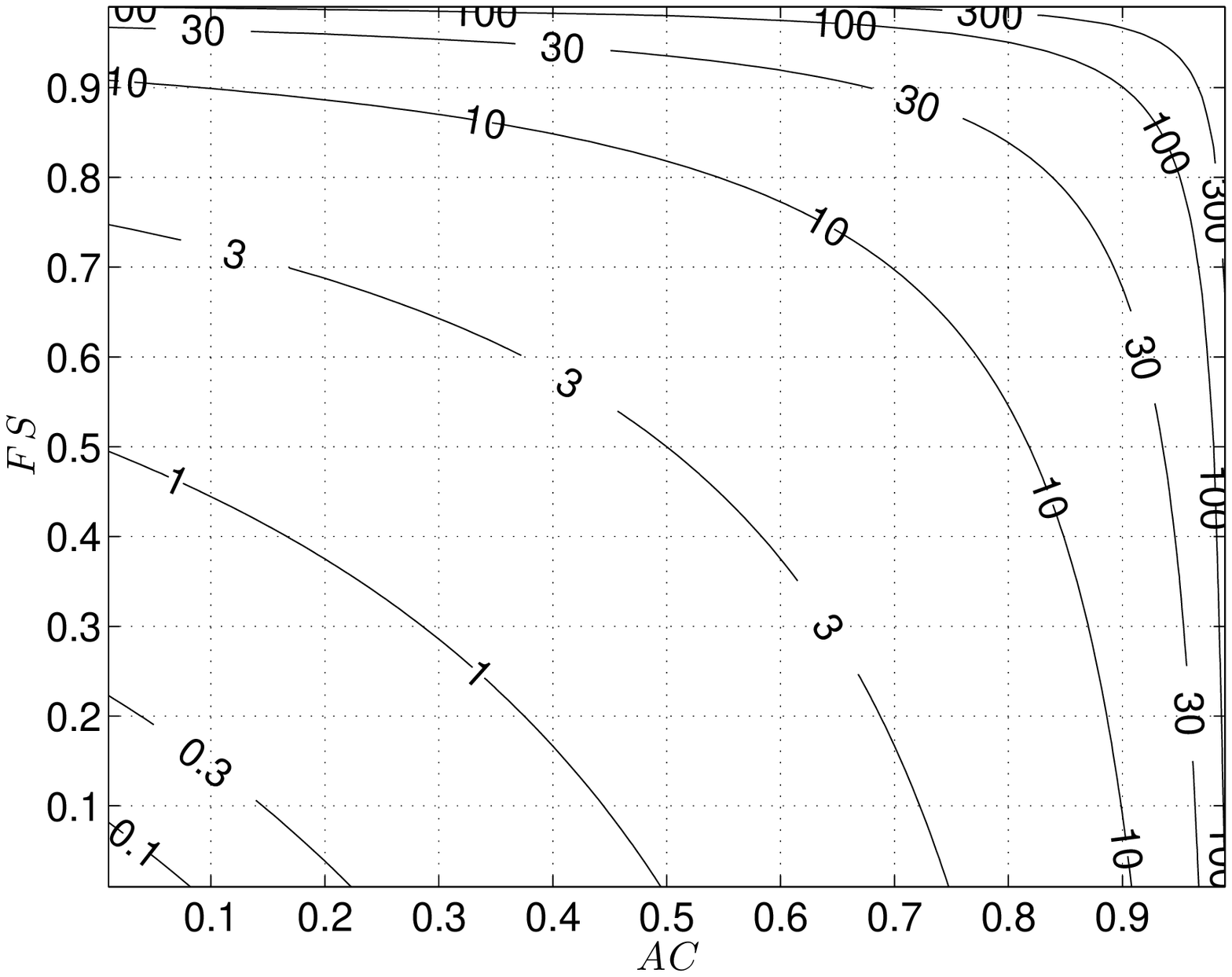}
      \caption{Contours of the inertial range $\log_{10}{\left(\mathcal{A}_{mT}\right)}$ for different mixtures of fluctuations determined by the fractions of Alfv\'{e}n to total energy ($AC$) and fast to total compressible energy ($FS$). }\label{MVA_contour}
\end{figure}

The assumptions that only slow modes with $k_\parallel \gg k_\perp$ are weakly damped and approximately parallel fast modes dominate the inertial range fast mode $\mathcal{A}_m$ lead to asymptotically small and large values of the slow and fast mode $\mathcal{A}_m$s, respectively. Due to the asymptotically large and small values of the component $\mathcal{A}_m$s, we can estimate the value of $\mathcal{A}_{mT}$ in the inertial range provided $AC \neq 1$ and $FS \neq 1$ by using the $\mathcal{A}_m$ of each mode from Figure \ref{fig:betas} to estimate the compressibilities, which have only a very weak $\beta_i$ dependence: $C_{\perp A} \simeq 1$, $C_{\parallel A} \ll 1$, $C_{\perp F} \simeq 1$, $C_{\parallel F} \ll 1$, $C_{\perp S} \ll 1$, and $C_{\parallel S} \simeq 1$. Therefore, in the inertial range,

\begin{equation}\label{eq:MVA_inertial}
\mathcal{A}_{mT} \simeq \frac{AC + (1-AC) FS}{(1-AC)(1-FS)}.
\end{equation}
Note that whether the slow modes are in fact propagating slow modes or PBSs does not effect the estimate because $C_{\perp S}$ is asymptotically small in either case. Therefore, the inertial range $\mathcal{A}_m$ depends only on the ratios of Alfv\'{e}n to total energy and the fast to total compressible energy ratio and is unable to differentiate between propagating slow modes and PBSs. Note that for observed average solar wind values of $AC \simeq 0.9$ and $FS \lesssim 0.1$ \citep{Howes:2011a,Klein:2012}, the $\mathcal{A}_m$ can be further reduced to $\mathcal{A}_{mT} \simeq AC / (1-AC)$. We take as fiducial solar wind values $AC = 0.9$ and $FS = 0.1$, or equivalently $90\%$ Alfv\'{e}n, $9\%$ slow, and $1\%$ fast wave energy. Although we take $FS = 0.1$, there is very little quantitative difference between $FS = 0.1$ and $0$ (see Figure \ref{MVA_contour}).

\begin{figure}[t]
		\includegraphics[width=\linewidth]{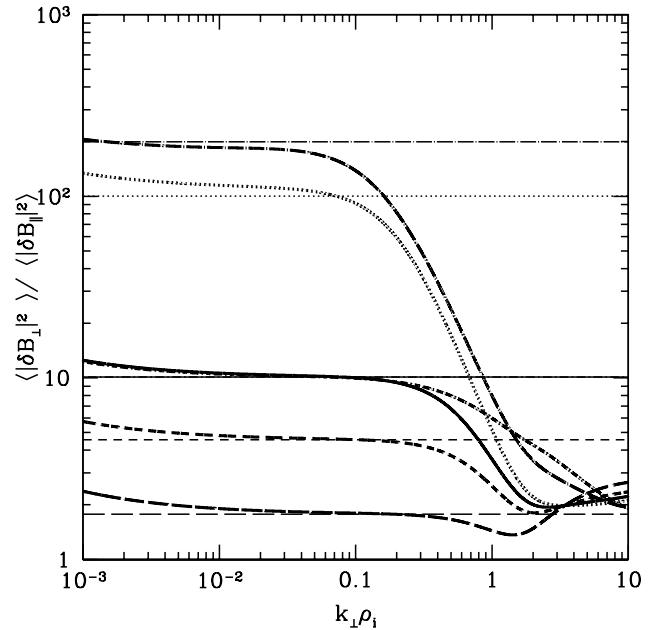}
      \caption{The sum of the Vlasov-Maxwell Alfv\'{e}n, fast, and slow modes for $\beta_i =1$ (thick) and the inertial range estimate given by equation~\eqref{eq:MVA_inertial} (thin) for different combinations of the three roots. The different line types represent Alfv\'{e}n to total energy ($AC$) and fast to total compressible energy ($FS$) fractions: $(AC,FS) = (0.9,0.1)$, $(0.99,0.1)$, $(0.8,0.1)$, $(0.6,0.1)$, $(0.1,0.9)$, and $(0.5,0.99)$ are solid, dotted, short dashed, long dashed, short dash-dotted, and long dash-dotted.}\label{fig:inertial_sum}
\end{figure}

Plotted in green in Figure \ref{fig:betas} is a single mixture of the three modes representing fiducial solar wind values of $AC = 0.9$ and $FS = 0.1$.  Clearly, the $\mathcal{A}_m$ has virtually no $\beta_i$ or wavenumber dependence in the inertial range, as expected from the above analysis. To explore the dependence of the $\mathcal{A}_m$ on fluctuation composition, we plot in Figure \ref{MVA_contour} the $\mathcal{A}_{mT}$ given by equation~\eqref{eq:MVA_inertial}. To confirm the quality of the estimate given by equation~\eqref{eq:MVA_inertial}, we plot in Figure \ref{fig:inertial_sum} the $\mathcal{A}_{mT}$ given by equation~\eqref{eq:MVA_full} (thick) and the estimate given by equation~\eqref{eq:MVA_inertial} (thin). The Figure demonstrates that equation~\eqref{eq:MVA_inertial} provides a good estimate for the total $\mathcal{A}_m$ in the inertial range. 

Also highlighted by the Figure is the degeneracy of the $\mathcal{A}_m$, since different mixtures of modes can replicate nearly identical $\mathcal{A}_m$s in the inertial range. This implies that the $\mathcal{A}_m$ when used alone is not a good metric for differentiating modes in the inertial range; however, the $\mathcal{A}_m$ can be a useful secondary metric. For instance, when used together with density-parallel magnetic field correlations to identify the proportion of fast to slow wave energy, the $\mathcal{A}_m$ can provide an estimate of the Alfv\'{e}n to compressible energy proportion, thereby quantifying the total population of the solar wind. 

\subsection{Transition and Dissipation Range}

The behaviour of the $\mathcal{A}_m$ in the transition between the inertial and dissipation ranges and in the dissipation range is markedly different from the inertial range. Unlike the inertial range, where the asymptotically large and small values of the Alfv\'{e}n, fast, and slow waves leads to an $\mathcal{A}_m$ that is controlled by the mode fractions $AC$ and $FS$, the KAW dominates the dissipation range $\mathcal{A}_m$  for typical solar wind parameters (see Figure \ref{fig:betas} for $k_\perp \rho_i > 1$, where the $\mathcal{A}_m$ for the mixture closely follows the $\mathcal{A}_m$ for KAWs).

The strong $\beta_i$ and $T_i / T_e$ dependence of the KAW $\mathcal{A}_m$ implies that for certain values, namely $\beta_i \gtrsim 1$, the whistler and KAW dissipation range $\mathcal{A}_m$ become approximately degenerate (see Figure \ref{fig:betas} for $k_\perp \rho_i > 1$). However, the behaviour through the transition range for the Alfv\'{e}n to KAW and fast to whistler mode transitions differs considerably. For example, the curves in Figure \ref{fig:inertial_sum} with $(AC,FS) = (0.9, 0.1)$ (solid black) and $(0.1, 0.9)$ (short dash-dotted black) are nearly identical in the inertial and dissipation ranges, but exhibit very different behaviour across the transition range. For this reason, the $\mathcal{A}_m$ is best presented as a function of wavenumber rather than as a quantity averaged over a band of wavenumbers, as has often been done in the literature \citep{Smith:2006a,Hamilton:2008,Smith:2012}.

\section{Measuring $\mathcal{A}_m$}\label{sec:measure}

The conventional method for calculating the $\mathcal{A}_m$ is as defined in \S \ref{sec:intro}; however, different methods of averaging can be performed. We here consider two physically motivated alternatives for measuring the $\mathcal{A}_m$ and compare them to the linear prediction. We also explore the validity of using linear theory to describe the $\mathcal{A}_m$ in a fully nonlinear turbulent situation.

\subsection{Synthetic Solar Wind Measurements}\label{sec:sw_model}

In the solar wind, the mean magnetic field direction is constantly changing and can sweep through a wide range of angles over a period of tens of minutes. As such, defining $\delta B_\parallel = \bB_0 \cdot \delta \bB$ becomes questionable since $\bB_0$ is averaged over long, global, periods relative to the rapidly fluctuating field---this is especially problematic when measuring in the dissipation range when $\delta B_\parallel$ is even smaller and more rapidly fluctuating. The poor definition of parallel and perpendicular in the global analysis will pollute the parallel energy with perpendicular energy, thereby decreasing the measured solar wind $\mathcal{A}_m$. Therefore, a local analysis employing a local mean magnetic field \citep{Cho:2000,Maron:2001,Horbury:2008,Podesta:2009a} must be employed to accurately measure $\delta B_\parallel$. The local analysis has the added advantage of being able to differentiate between different $\theta_{VB}$ measurements, where $\theta_{VB}$ is the angle between the mean magnetic field and the solar wind flow velocity. This can be helpful because the $\mathcal{A}_m$ will have different inertial/dissipation range breakpoints for KAWs and whistlers when plotted against $k_\parallel$ or $k_\perp$. 

The conventional $\mathcal{A}_m$ is constructed by calculating separately the perpendicular and parallel fluctuating magnetic energies and finding their quotient, 

\begin{equation}\label{eq:conv_MVA}
\mathcal{A}_m = \frac{\langle|\delta B_\perp|^2\rangle}{\langle|\delta B_\parallel|^2\rangle}.
\end{equation} 
However, a similar measure could be constructed from the normalized perpendicular and parallel energies, $\langle |\delta B_\perp|^2 / |\delta \bB|^2 \rangle$ and $\langle |\delta B_\parallel|^2 / |\delta \bB|^2 \rangle$,
\begin{equation}\label{eq:alt_1}
\mathcal{A}_m = \frac{\langle |\delta B_\perp|^2 / |\delta \bB|^2 \rangle}{\langle |\delta B_\parallel|^2 / |\delta \bB|^2 \rangle}.
\end{equation}
This measure has the physically motivated advantage of avoiding possible small denominators due to small $\delta B_\parallel$. Another possible measure of the $\mathcal{A}_m$ is
 \begin{equation}\label{eq:alt_2}
\mathcal{A}_m = \langle |\delta B_\perp|^2 / |\delta B_\parallel|^2\rangle.
\end{equation} 
This method has the conceptual advantage of maximizing any local effect of the mean magnetic field since it averages the $\mathcal{A}_m$ at each point rather than separately averaging the energies. However, the expression $\langle |\delta B_\perp|^2 / |\delta B_\parallel|^2\rangle$ will be dominated by those terms with small denominators, which could make it an unphysical measure of the $\mathcal{A}_m$

To explore the effect different averaging procedures have on the $\mathcal{A}_m$, we employ synthetic spacecraft data. The details for constructing the synthetic spacecraft data are discussed in \citet{Klein:2012}, so we here only detail the relevant parameters. A three-dimensional box is populated with a spectrum of all three wave modes, where the proportion of each mode is given by $AC = 0.9$ and $FS = 0.1$, corresponding to $90\%$ Alfv\'{e}n, $9\%$ slow, and $1\%$ fast wave energy. The Alfv\'{e}n and slow modes satisfy critical balance, with all modes less than the critical balance envelope equally populated. The fast modes are isotropically populated from $\theta_{kB} = 5 - 85^\circ$, with equal energy at each angle. To represent solar wind turbulence, the phase of the wave modes is randomized. The data is sampled by advecting the turbulence past a stationary "spacecraft" with velocity $v = 10 v_A$ at a fixed angle with respect to the mean magnetic field and fixed sampling rate. 

\begin{figure}[t]
		\includegraphics[width=\linewidth]{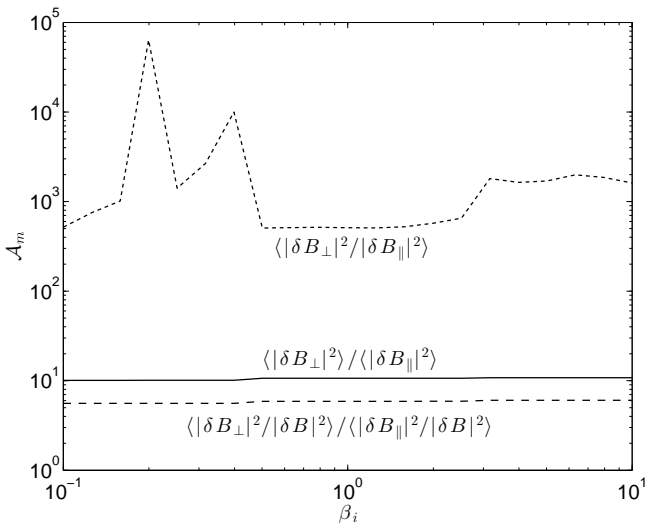}
      \caption{Synthetic spacecraft data populated with $90\%$ Alfv\'{e}n, $9\%$ slow, and $1\%$ fast wave energy averaged over $k_\perp \rho_i \in [0.003, 0.03]$. The three different lines represent different averaging procedures for calculating the $\mathcal{A}_m$.}\label{fig:synth}
\end{figure}

Figure \ref{fig:synth} presents the three different methods for calculating the $\mathcal{A}_m$. The $\mathcal{A}_m$ is averaged across the interval $k_\perp \rho_i \in [0.003, 0.03]$ with $\theta_{vB} = 85^\circ$, where $\theta_{vB}$ is the sampling angle analogous to that in solar wind. Sampling angle, $\theta_{vB}$, was varied from $5 - 85^\circ$ and found to have no significant effect on the $\mathcal{A}_m$ in the inertial or dissipation range, but sampling angle does alter the behaviour in the transition region. 

For $AC = 0.9$ and $FS = 0.1$, the predicted $\mathcal{A}_m$ from linear theory is $\mathcal{A}_{mT} \simeq 10$. From the Figure, it is clear that the conventional definition of the $\mathcal{A}_m$, equation ~\eqref{eq:conv_MVA}, (solid) computed from a spectrum of wave modes agrees best with linear theory and is the one we will continue to employ. The averaged anisotropy, equation ~\eqref{eq:alt_1}, (dotted) approach is a poor measure of the $\mathcal{A}_m$ because at any given point in space, the superposition of a collection of wave modes can lead to anomalously small values of $\delta B_\parallel$ due to cancellation. While the definition based upon normalized energies, equation ~\eqref{eq:alt_2}, (dashed) differs from the linear prediction, it captures the correct qualitative behaviour and may be a safer measure to use in some circumstances.

\subsection{Nonlinear Simulation Measurements}\label{sec:sim}

Here we present results from a collection of fully nonlinear gyrokinetic turbulence simulations performed with $\beta_i = 0.1$ and $1$, $T_i /T_e = 1$, and a realistic mass ratio, $m_i / m_e = 1836$. The simulations were performed using the Astrophysical Gyrokinetics Code, AstroGK \citep{Numata:2010}. All runs presented herein are driven with an oscillating Langevin antenna coupled to the parallel vector potential at the simulation domain scale \citep{TenBarge:2012b}. Relevant parameters for the simulations are given in table \ref{tab:AGK}, where $\epsilon = \rho_i / L_0 \ll 1$ is the gyrokinetic expansion parameter, $A_0$ is the antenna amplitude, and $\nu_s$ is the collision frequency of species $s$. The expansion parameter sets the parallel simulation domain elongation and is determined by assuming critical balance with an outer-scale $k_i \rho_i = 10^{-4}$: $\epsilon = \hat{k}_i^{1/3} \hat{k}_{\perp0}^{2/3}  (1 + \hat{k}_{\perp0}^{5/3}) / (1 + \hat{k}_{\perp0}^2)$, where $\hat{k} = k \rho_i$ and subscript naught indicates simulation domain scale quantities. The antenna amplitude is chosen to satisfy critical balance at the domain scale so that the simulations all represent critically balanced, strong turbulence.

\begin{table*}[t]
\caption{Parameters for AstroGK simulations used in Figures \ref{fig:beta1_spec}, \ref{fig:b01_agk_all}, and \ref{fig:b1_agk_all}.}\label{tab:AGK}
\begin{center}
\begin{tabular}{|c||c|c|c|c|c|c|}
\hline \hline
        Run & $k_\perp \rho_i$ & $k_\parallel \rho_i / \epsilon$ & $\epsilon$ & $A_0 / \epsilon \rho_i B_0$ & $\nu_i / \epsilon \Omega_i$ & $\nu_e / \epsilon \Omega_i$ \\
\hline
$\beta0.1$I & $[0.05,1.05]$  & $[1,64]$ & 0.006 & 500 & 0.001 & 0.03\\
\hline
$\beta0.1$T & $[0.2,8.4]$ & $[1,64]$ & 0.016 & 35 & 0.003 & 0.05\\
\hline
$\beta0.1$D & $[1,42]$  & $[1,64]$ & 0.046 & 1 & 0.02 & 0.4\\
\hline
$\beta1$I & $[0.05,1.05]$  & $[1,64]$ & 0.006 & 500 & 0.001 & 0.0005 \\
\hline
$\beta1$T & $[0.2,4.2]$  & $[1,64]$ & 0.016 & 30 & 0.008 & 0.03 \\
\hline
$\beta1$D & $[1,42]$  & $[1,64]$ & 0.046 & 1 & 0.04 & 0.5\\
\hline
$\beta1$ED  & $[5,105]$  & $[1,16]$ & 0.081 & 0.2 & 0.2 & 0.5\\ 
\hline
\end{tabular}
\end{center}
\end{table*}

\begin{figure}[t]
		\includegraphics[width=\linewidth]{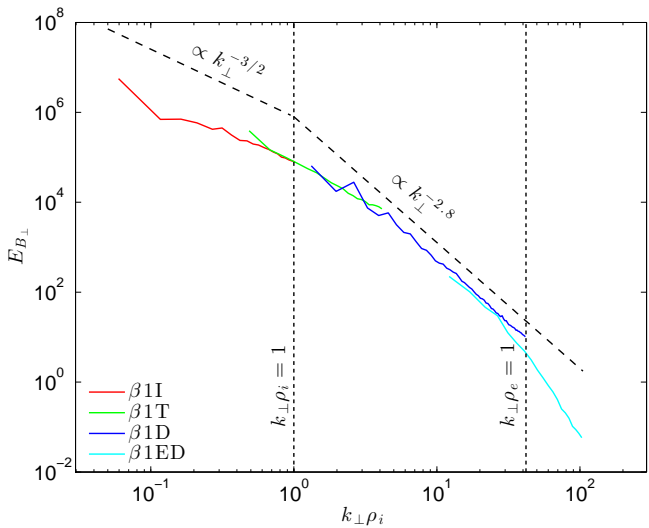}
      \caption{(Color online) Composite one-dimensional perpendicular magnetic energy spectrum composed of four overlapping $\beta_i = 1$ AstroGK turbulence simulations spanning from the inertial range to deep into the dissipation range.}\label{fig:beta1_spec}
\end{figure}

An example instantaneous one-dimensional perpendicular magnetic energy spectrum composed by overlaying the four $\beta_i = 1$ simulations is plotted in Figure \ref{fig:beta1_spec}---the spectrum is produced via conventional Fourier analysis techniques. The inertial range simulation (red) has a spectral index $\sim -3/2$, which steepens (green) to $-2.8$ (blue) before rolling off exponentially (cyan) as the electron collisionless dissipation becomes increasingly strong. The dissipation range simulations $\beta1$D and $\beta1$ED have been explored in detail \citep{Howes:2011b,TenBarge:2012c}. The full spectrum agrees well with recent solar wind observations \citep{Alexandrova:2011}.

\begin{figure*}[t]
\begin{center}	
	\subfloat[]{\label{fig:b01_agk_all}\includegraphics[width=0.5\textwidth]{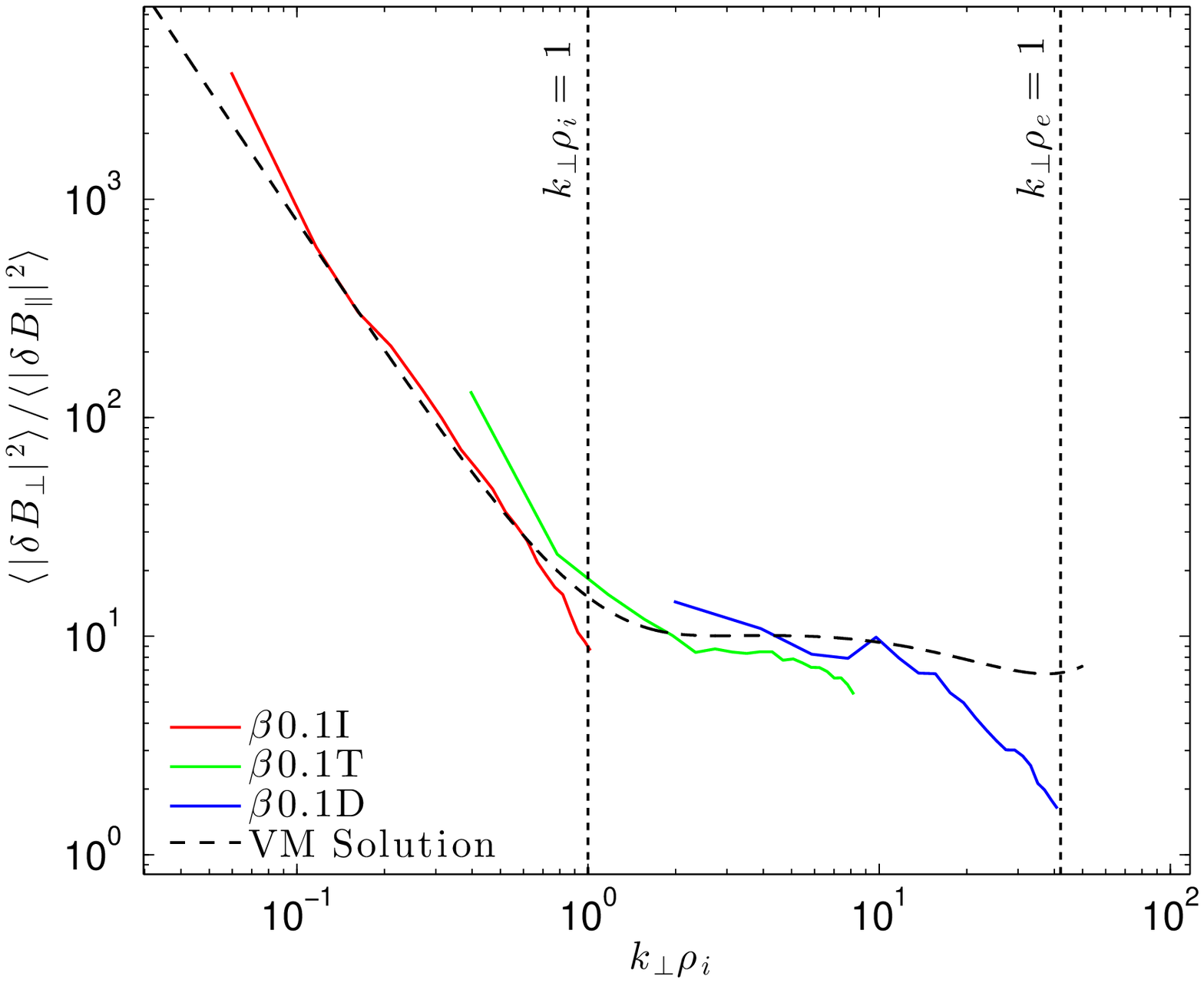}}                
 	\subfloat[]{\label{fig:b1_agk_all}\includegraphics[width=0.5\textwidth]{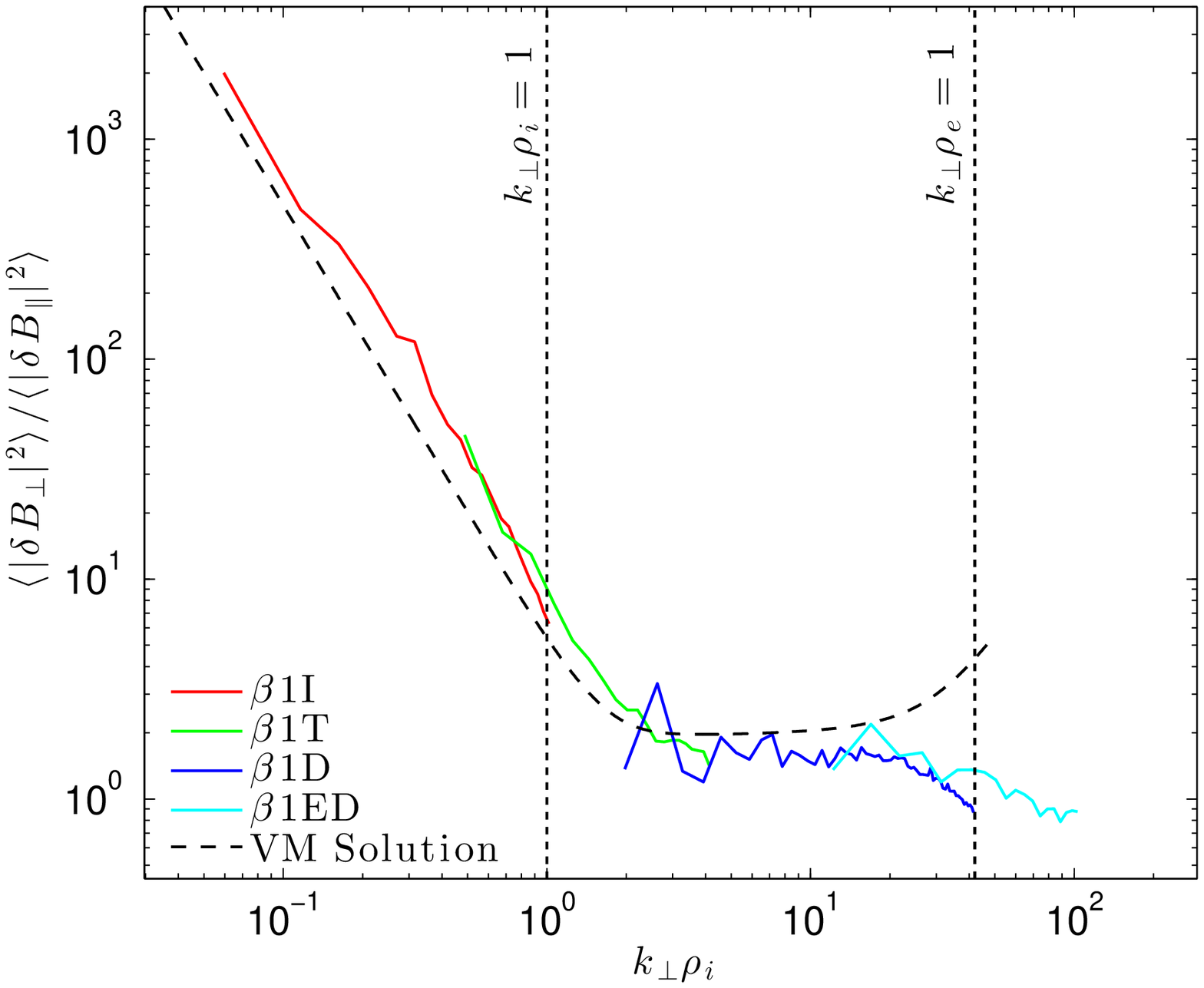}}
\end{center}
  \caption{(Color online) A comparison of the linear Vlasov-Maxwell solution (dashed) to overlaid nonlinear gyrokinetic simulations of Alfv\'{e}nic turbulence spanning from the inertial range to deep into the dissipation range for (a) $\beta_i = 0.1$ and (b) $\beta_i = 1$.}
\end{figure*}

Figures \ref{fig:b01_agk_all} and \ref{fig:b1_agk_all} overlay the $\mathcal{A}_m$ for the $\beta_i = 0.1$ and $1$ AstroGK simulations. The $\mathcal{A}_m$ is produced by computing, via conventional Fourier techniques, the perpendicular and parallel energy with respect to the global mean magnetic field. Due to the restrictions of gyrokinetics \citep{Howes:2006} and the method of driving, the simulations are almost purely populated by Alfv\'{e}n / KAWs. Therefore, we plot as a dashed line in each Figure the VM linear solution for the Alfv\'{e}n root. The agreement between the linear prediction and the nonlinear simulation is excellent up to the point of strong damping at $k_\perp \rho_i \sim 10$, where agreement with linear theory is expected to breakdown. The good agreement supports the validity of using linear theory to describe the $\mathcal{A}_m$ of fully nonlinear turbulence. 

Note that in constructing Figures \ref{fig:b01_agk_all} and \ref{fig:b1_agk_all} a global mean magnetic field has been used. This is justified, because in typical numerical simulations, the global mean field direction is fixed and $|\delta \bB| / B_0 \ll 1$. Although a local mean field can be defined in the same sense as in the solar wind, the global mean field provides a relatively accurate measure of the $\mathcal{A}_m$ in numerical simulations of turbulence with a fixed global mean field direction and $|\delta \bB| / B_0 \ll 1$.

\section{Solar Wind Measurements}\label{sec:solar}

We here discuss some of the recent solar wind analyses of the $\mathcal{A}_m$, present new solar wind measurements of the $\mathcal{A}_m$, and compare the new measurements to our predictions.

\subsection{Previous Results}\label{sec:previous}

\citet{Smith:2006a} presents the most comprehensive collection of the solar wind inertial range $\mathcal{A}_m$ made to date, incorporating 960 data intervals recorded by the ACE spacecraft. The data is divided into periods of open field lines and magnetic clouds \citep{Burlaga:1981}. They find that the $\mathcal{A}_m$ is proportional to a power of $\beta_i$ or $|\delta \bB| / B_0$ but are unable to identify which is the source of the functional relationship since $\beta_i$ and $|\delta \bB|$ are themselves positively correlated in the solar wind \citep{Grappin:1990}. They follow the identification of the $\mathcal{A}_m$ relationship with a discussion of possible sources for the relationship based on a variety of considerations, some of which contradict the discussion in \S \ref{sec:waves} of this paper. \citet{Smith:2006a} do not consider the superposition of modes that has been shown to exist in the solar wind, neglect the contribution of slow modes due to the belief that they are completely damped outside of local excitation, consider Alfv\'{e}n waves to be completely incompressible ($\mathcal{A}_m = \infty$), consider PBSs but assert that the $\beta_i$ dependence of $\delta B_\parallel$ of PBSs could contribute the the $\beta_i$ dependent $\mathcal{A}_m$ despite $\mathcal{A}_{m{PBS}} = 0$ for all $\beta_i$.

The findings of \citet{Smith:2006a} seem to be in direct contradiction to the conclusions drawn in \S \ref{sec:MVA_inertial}; however, there are possible reasons for the apparent discrepancy. As noted in \citet{Smith:2006a}, the proton temperature and fluctuating magnetic field are positively correlated. It has also been shown that the proton temperature and bulk solar wind speed are positively correlated \citep{Burlaga:1973,Newbury:1998}. Therefore, the measured $\mathcal{A}_m$ might depend on any or a combination of $\beta_i$, $|\delta \bB|$, or $v_{SW}$. If the underlying dependence is actually on $v_{SW}$, then the observed variation of the $\mathcal{A}_m$ could be from different types of solar wind launched from different regions of the sun having a variable population of Alfv\'{e}n to compressible components. For instance, slower wind may have evolved to a more Alfv\'{e}nic state, because fast modes would have had time to be dissipated in shocks and slow modes could have collisionlessly damped, leaving a primarily Alfv\'{e}nic and PBS population. Since no measure other than $\mathcal{A}_m$ binned by $\beta_i$ or $|\delta \bB| / B_0$ was used in \citep{Smith:2006a}, it is difficult to draw a firm conclusion.

\citet{Hamilton:2008} extend the study of the same dataset employed in \citet{Smith:2006a} to the dissipation range, defined to be $0.3$~Hz~$ \le f \le 0.8$~Hz . They find that dissipation range fluctuations are less anisotropic than those in the inertial range. Although not discussed in \citet{Hamilton:2008}, this result is consistent with the mixtures of wave modes discussed in this paper. The linear predictions of KAW $\mathcal{A}_m$ presented in Figure \ref{fig:kaw_beta_dep} for $T_i /T_e = 1$ pass through the core of the measured $\mathcal{A}_m$ for open magnetic field data for the measured $\mathcal{A}_m$ in the dissipation range presented in the lower panel of Figure 8 of \citet{Hamilton:2008}. Further, the KAW $\mathcal{A}_m$ for $T_i/T_e = 0.1$ and $T_i/T_e =1$ bound the core of the low $\beta_i$ magnetic cloud data for the measured dissipation range $\mathcal{A}_m$ in \citet{Hamilton:2008}. The temperature ratio is not provided for the dataset in \citet{Hamilton:2008}; however, magnetic clouds are typically characterized by low proton temperatures \citep{Osherovich:1993,Richardson:1997}, so $T_i / T_e < 1$ for the cloud data is expected. \citet{Gary:2009} also noted that the $\beta_i > 0.1$ portion of the same dataset is well fit by the KAW solution; however, the temperature ratio dependence was not explored. \citet{Gary:2009} suggest the shallow slope at low $\beta_i$ might be suggestive of a whistler component. Another possible explanation for some of the very low values of the dissipation range $\mathcal{A}_m$ in \citet{Hamilton:2008} is the use of a global measurement of the magnetic field, which will tend to decrease the measured $\mathcal{A}_m$, especially in the dissipation range. Also, due to the limited high frequency information available from ACE measurements $f_{Nyquist} \simeq 2$~Hz \citep{Smith:1998}, the region defined as the dissipation range corresponds to the transition region between the inertial and dissipation ranges.

\citet{Smith:2012} return again to the same ACE dataset but focus on $\beta_i > 1$ intervals and augment the set with 29 additional high $\beta_i$ ACE measurements. As such, this dataset suffers from the same high frequency limitations noted above for the \citet{Hamilton:2008} analysis. \citet{Smith:2012} suggest that the measured dissipation range $\mathcal{A}_m$ is consistent with KAWs, but the inertial range $\mathcal{A}_m$ cannot be explained by a purely Alfv\'{e}nic population. This observation is partially used to conclude that the Alfv\'{e}n/KAW model cannot explain solar wind measurements. However, their $\mathcal{A}_m$ measurements agree well with a solar wind population dominated by Alfv\'{e}n/KAWs with a small component of slow waves. Therefore, we find no contradiction between the ACE $\mathcal{A}_m$ measurements and the Alfv\'{e}n mode dominant $\mathcal{A}_m$ model constructed herein.

Using STEREO measurements, \citet{He:2012} employ a variant of the $\mathcal{A}_m$, $\delta B_\parallel / \delta B_\perp$, as a secondary metric to the magnetic helicity to differentiate between KAWs and whistlers, which both have right-handed helicity for oblique propagation\citep{Howes:2010a}. However, they consider fast modes propagating at fixed $\theta_{kB} = 60^\circ$, $80^\circ$, and $89^\circ$, and only the $\theta_{kB} = 60^\circ$ root connects to an oblique whistler. The other two oblique roots connect to IBWs, which is made clear in the upper right panel of Figure 3 of \citet{He:2012}, where the dispersion relation for these two roots do not extend above $\omega / \Omega_i = 1$. Due to this mischaracterization of the roots, they mistakenly find the oblique whistler root $\mathcal{A}_m \ll 1$. As noted in \S \ref{sec:fastmode}, the whistler (fast mode for $k_\parallel d_i > 1$) $\mathcal{A}_m > 1$ for all propagation angles and $\beta_i$ and is thus of the same orientation as KAW and does not provide a useful secondary metric to the helicity.

Using Cluster measurements, \citet{Salem:2012} use magnetic compressibility, which is related to the $\mathcal{A}_m$ in equation~\eqref{eq:para_comp}, as a secondary metric to the ratio of the electric to magnetic field, $|\delta \bE| / |\delta \bB|$, which is degenerate between KAWs and whistlers. Although the dissipation range asymptotic behaviour of the fixed propagation angle KAWs and the whistlers at certain angles is similar, the transition from the inertial range to dissipation range for the two modes is different, and it is the transition region that is used to differentiate the two modes. \citet{Salem:2012} find the Cluster data to be most consistent with a spectrum of KAWs.

Although much of the work does not directly address the $\mathcal{A}_m$, some recent solar wind analyses of the magnetic field have moved toward three dimensions \citep{Chen:2011,Podesta:2012,Wicks:2012}. Presenting the full three dimensional structure of the magnetic field is of obvious potential benefit; however, there are complications due to geometrical and sampling effects of the fluctuations being advected past the spacecraft \citep{Turner:2011,Wicks:2012}. To elucidate the complications, we follow \citet{Turner:2011} and choose a coordinate system with $B_0$ in the $z$ direction and $V_{sw}$ in the $y$ direction. Under Taylor's hypothesis, the measured components of magnetic energy in the perpendicular plane are then
\[
E_{x,y} (\omega)= \frac{1}{8\pi} \int d^3 \bk |\delta \bB_{x,y}|^2 \delta(\omega - \bk \cdot \bV_{sw} ).
\] 
If a scaling between $k_\parallel$ and $k_\perp$ is assumed, then
\[
E_x(\omega) = \int d^2\bk_\perp E_{2D}(\bk_\perp)\delta(\omega - k_y V_{sw}) \cos^2{(\phi)}
\]
\[
E_y(\omega) = \int d^2\bk_\perp E_{2D}(\bk_\perp)\delta(\omega - k_y V_{sw}) \sin^2{(\phi)},
\]
where $E_{2D}(\bk_\perp) = C k_\perp^{-\xi-1}$ is the two-dimensional spectrum, $\xi$ is the one-dimensional scaling exponent, and $\tan{(\phi)} = k_y/k_x$. \citet{Turner:2011} find that $E_y/E_x = 1 / \xi$, implying a nonaxisymmetric energy distribution in the perpendicular plane. This complication is obviated by measuring $\mathcal{A}_m$ in the standard way, since 
\[
\begin{split}
&E_\perp(\omega)  = E_x(\omega) + E_y(\omega) = \\
&\int d^2\bk_\perp E_{2D}(\bk_\perp)\delta(\omega - k_y V_{sw})
\end{split}
\]
avoids the angular sampling issue altogether.

\subsection{New Measurements}\label{sec:new}

To compare the analysis method outlined in \S \ref{sec:waves} and \ref{sec:combos} to solar wind measurements, we choose a 5 day interval, 2008 Feb 12 06:00 to Feb 17 06:00, of magnetic field data captured by the Stereo A spacecraft. This interval is one of 20 similar intervals analysed by \citet{Podesta:2012}, where details concerning the data selection and analysis can be found. So, we here summarize only the most pertinent aspects of the measurement and analysis. The interval was chosen because it represents an interval of high-speed wind, which typically satisfies the assumption that the solar wind flow direction is in the heliocentric radial direction, $\bR$. Also, the interval is sufficiently long to include a statistically large sample of periods during which the mean magnetic field, $\bB_0$, is nearly perpendicular to the radial direction, which in practice includes the range $84^\circ < \theta_{VB} < 96^\circ$. Focusing on these orthogonal periods is helpful because it facilitates easier comparison to theory since the measured wavenumber in the solar wind corresponds more closely to the perpendicular wavenumber used in linear theory. 

The magnetic field data was analysed using wavelet techniques to determine the local mean magnetic field, $\bB_0(\tau)$, at a given scale $\tau$ \citep{Horbury:2008,Podesta:2009a}. The parallel and perpendicular magnetic energies are then given by $\delta B_\parallel(\tau)^2 = | \hat{\bB}_0(\tau) \cdot \delta \bB(\tau) |^2$ and $\delta B_\perp(\tau)^2 = |\delta \bB(\tau)|^2 - \delta B_\parallel(\tau)^2$, where $\hat{\bB}_0(\tau) = \bB_0(\tau) / B_0(\tau)$.

The proton plasma parameters for the interval were supplied by the PLASTIC instrument \citep{Galvin:2008}. The relevant proton plasma parameters are $V_p = 657~km/s$, $n_p =2.6~cm^{-3}$, $T_p = 1.7 \times 10^5~K$, $\beta_p = 0.65$, where all quantities are averaged over the entire interval except $\beta_p$ whose median is taken because of its high variability. Electron thermal data is unavailable due to problems with the IMPACT Solar Wind Electron Analyzer aboard Stereo \citep{Fedorov:2011}, so we assume the electron temperature to be the average for fast wind streams found by \citet{Newbury:1998}, $T_e = 1.4 \times 10^5~K$. This assumption for the electrons implies $T_p / T_e \simeq 1$ for the data interval.

To convert spacecraft-frame frequency, $f_{sc}$, to wavenumber, we employ Taylor's hypothesis, which states that $2 \pi f_{sc} = \omega_{sc} = \omega_p + \bk \cdot \bV_{sw}$, where $\omega_p$ is the plasma rest-frame frequency. For oblique Alfv\'{e}nic and KAW fluctuations, $\omega_p \ll \bk \cdot V_{sw}$ \citep{Howes:2012a,Smith:2012}, and we can assume $2 \pi f_{sc} = k V_{sw}$. To compare to linear theory, we convert spacecraft frame frequency to wavenumber using $k \rho_i = 2 \pi \rho_i f_{sc} / V_{sw}$, where $\rho_i$ is computed from the average solar wind proton quantities for the measured interval.

\begin{figure}[t]
		\includegraphics[width=\linewidth]{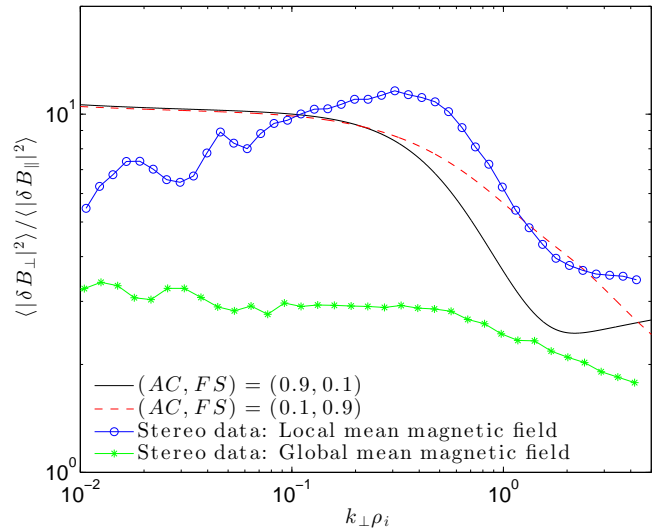}
      \caption{(Color online) A comparison between the measured Stereo A $\mathcal{A}_m$ data computed via a local wavelet analysis (blue) in the $84^\circ < \theta_{VB} < 96^\circ$ bin, the measured Stereo A $\mathcal{A}_m$ computed with a global, $1$~hr averaged, mean magnetic field (green), and the summed Vlasov-Maxwell $\mathcal{A}_m$ with $\beta_i = 0.65$ and $T_i / T_e = 1$ employing $(AC, FS) = (0.9, 0.1)$ (black) and $(AC, FS) = (0.1, 0.9)$ (red dashed).}\label{fig:stereo}
\end{figure}

In Figure \ref{fig:stereo}, we compare the measured $\mathcal{A}_m$ for this interval (blue) to the $\mathcal{A}_m$ from linear theory with $\beta_i = 0.65$ and $T_i / T_e = 1$ for two different mixtures of wave modes that reproduce the solar wind $\mathcal{A}_m$ in the inertial range: a mixture that is dominantly Alfv\'{e}nic (black) with $(AC, FS) = (0.9, 0.1)$ and a mixture that is dominantly fast mode (red dashed) with $(AC, FS) = (0.1, 0.9)$. Although both mixtures of linear wave modes reproduce well the behaviour in the inertial range, the fast mode dominant construction displays markedly different transition and dissipation range behaviour; whereas, the Alfv\'{e}n dominant mixture fits well the \textit{slope} of the transition range and the asymptotic value in the dissipation range. From this, we can conclude that the measured interval is most likely dominated by Alfv\'{e}n wave fluctuations at $k_\perp \rho_i \lesssim 1$ and KAW fluctuations for $k_\perp \rho_i \gtrsim 1$. 

We speculate that the positive slope in the measured inertial range $\mathcal{A}_m$ may be attributable to a decreasing fraction of slow wave energy due to collisionless damping as the turbulent cascade proceeds to kinetic scales. Also, the approximate factor of two difference between the predicted $\mathcal{A}_m$ from linear theory and the solar wind values in the transition and dissipation ranges are likely due to averaging $\beta_i$ and $\rho_i$ over the entire 5 day interval. The $\mathcal{A}_m$ of KAWs has a moderately strong $\beta_i$ dependence (see Figures \ref{fig:betas} and \ref{fig:kaw_beta_dep}). For the purposes of comparison, we use a median $\beta_i = 0.65$. However, the $\mathcal{A}_m$ of KAWs increases with decreasing $\beta_i$, so periods of low $\beta_i$ will tend to dominate the average $\mathcal{A}_m$ and shift it upward from the prediction based on a median value of $\beta_i = 0.65$. Similarly, we use a value for $\rho_i$ based on quantities averaged over 5 days of data to determine the relationship between frequency and wavenumber. The variance associated with the averaged $\rho_i$ could cause an apparent horizontal shift of the data in the transition and dissipation ranges.

For comparison, we also plot in Figure \ref{fig:stereo} the measured $\mathcal{A}_m$ for the Stereo A data computed via a more conventional global analysis (green). The $5$~day interval of data was segmented into $120$ $1$~hr intervals. In each $1$~hr segment, the mean magnetic field was calculated, the data rotated into mean field coordinates, and a Welch windowed FFT was applied to compute the parallel and perpendicular magnetic energies. The parallel and perpendicular magnetic energies were then averaged over the $120$ $1$~hr segments and binned into logarithmic wavenumber bins to smooth the spectra. The global $\mathcal{A}_m$ (green) was then calculated from the binned and averaged magnetic field data. As expected, this method is inappropriate to compare to linear theory and produces a significantly reduced $\mathcal{A}_m$ compared to the wavelet analysis because the parallel magnetic energy is polluted with perpendicular energy, as discussed in \S \ref{sec:sw_model}.

This analysis of solar wind data highlights the importance of performing a local analysis and the weaknesses and strengths of the $\mathcal{A}_m$ for determining the solar wind fluctuation composition: The $\mathcal{A}_m$ is a poor tool to use in the inertial range due to degeneracies of different wave mode mixtures. However, if the inertial range $\mathcal{A}_m$ is augmented with a second measure that identifies the the fast to total compressible energy fraction ($FS$), such as the density-parallel magnetic field correlation, the $\mathcal{A}_m$ can provide a measure of the Alfv\'{e}n to total energy fraction ($AC$). Also, the slope of the $\mathcal{A}_m$ transition range and the asymptotic value in the dissipation range are useful for identifying the solar wind fluctuation composition.


\section{Conclusions}\label{sec:conc}

We have developed a framework for interpreting the measured solar wind magnetic variance anisotropy ($\mathcal{A}_m$) and comparing the measurements to linear theory. In \S \ref{sec:waves} we reviewed the linear properties of the collisionless counterparts to the three MHD wave modes, which are expected to be the primary constituents of the solar wind in those relevant regions of wavenumber space spanning the inertial and dissipation ranges where collisionless damping rate is small compared to the rate nonlinear energy transfer. The $\mathcal{A}_m$ of each wave mode for a range of plasma betas, $\beta_i$, is shown in Figure \ref{fig:betas} versus perpendicular wavenumber.

In \S \ref{sec:combos} we examined how superpositions of the three wave modes affects the $\mathcal{A}_m$. Due to the asymptotically large $\mathcal{A}_m$ of the Alfv\'{e}n and fast modes and asymptotically small $\mathcal{A}_m$ of the slow mode in the inertial range, the inertial range $\mathcal{A}_m$ is dictated by the fraction of each mode rather than the individual behaviour of each mode and has little $\beta_i$ dependence. Note that linear pressure balanced structures (PBSs) are equivalent to $k_\parallel = 0$, non-propagating slow modes, and are thus classified as slow modes. An estimate of the inertial range $\mathcal{A}_m$ is given by equation ~\eqref{eq:MVA_inertial}, where the $\mathcal{A}_m$ is determined by the fraction of Alfv\'{e}n to total (Alfv\'{e}n plus fast plus slow) energy and the fraction of fast to total compressible (fast plus slow) energy. For $\beta_i \gtrsim 1$, the dissipation range value of the $\mathcal{A}_m$ for kinetic Alfv\'{e}n waves (KAWs) and whistler waves is approximately degenerate; however, the behaviour of the $\mathcal{A}_m$ through the transition to the dissipation range differs in form for KAWs and whistlers. The degeneracy in the dissipation range highlights the value of displaying the $\mathcal{A}_m$ as a function of wavenumber.



In \S \ref{sec:sim} we employ a suite of fully nonlinear gyrokinetic turbulence simulations that span from the inertial range to deep into the dissipation range for $\beta_i = 0.1$ and $1$ to demonstrate that the predictions of the $\mathcal{A}_m$ from linear theory are applicable to nonlinear turbulence. The comparison between the linear prediction for the $\mathcal{A}_m$ of Alfv\'{e}n waves and the $\mathcal{A}_m$ measured in the nonlinear turbulence simulation are presented in Figure 9, where excellent agreement is seen up to kinetic scales where electron dissipation becomes significant, $k_\perp \rho_i \simeq 10$.

Some of the recent solar wind measurements of the $\mathcal{A}_m$ and their interpretations are presented in \S \ref{sec:previous}. Since the $\mathcal{A}_m$ of previous solar wind studies typically bins the $\mathcal{A}_m$ over a band of wavenumbers and computes the $\mathcal{A}_m$ via a global analysis, the data is difficult to interpret and compare to linear theory; however, the $\mathcal{A}_m$ observed in previous studies is mostly consistent with solar wind composed primarily of Alfv\'{e}nic fluctuations and conforms well to the theory outlined in this paper.


\S \ref{sec:new} presented a measurement of the solar wind $\mathcal{A}_m$ from Stereo A performed using the methods outlined in \S \ref{sec:sw_model}, computed locally via wavelet analysis techniques, and plotted as a function of perpendicular wavenumber. We find excellent agreement between linear theory and the local solar wind measurement, which suggests that this interval of solar wind data is composed of approximately $90\%$ Alfv\'{e}n, $10\%$ slow, and negligible fast wave energy across the sampled range of scales spanning the inertial range to the beginning of the dissipation range. Here, slow wave energy encompasses both PBSs and propagating slow modes because the $\mathcal{A}_m$ alone cannot differentiate between the two. This analysis highlights the advantages of computing the $\mathcal{A}_m$ locally with a wavelet analysis and plotting the $\mathcal{A}_m$ as a function of wavenumber rather than binning across wavenumber bands because the transition range breaks the degeneracy of the inertial range. Alternatively, a second measure such as the density-parallel magnetic field correlation can be used to break the degeneracy and ascertain a value for fast to total compressible energy fraction ($FS$). The $\mathcal{A}_m$ can then be used to measure the Alfv\'{e}n to total energy energy fraction ($AC$).

In summary, we have outlined the salient features of linear theory as they relate to the magnetic variance anisotropy, provided a procedure for producing and comparing solar wind measurements to predictions from linear theory, demonstrated the validity of this technique via both nonlinear turbulence simulations and solar wind measurements, and found previous studies of the solar wind magnetic variance anisotropy and new solar wind data to be consistent with a dominantly Alfv\'{e}nic population in the inertial \textit{and} dissipation ranges.

\section*{Acknowledgements}
This work was supported by NASA grant NNX10AC91G and NSF CAREER grant
AGS-1054061. This research used resources of the Oak Ridge Leadership
Computing Facility at the Oak Ridge National Laboratory, supported by
the Office of Science of the U.S. Department of Energy under Contract
No.  DE-AC05-00OR22725.  This research was supported by an allocation
of advanced computing resources provided by the National Science
Foundation, partly performed on Kraken at the National Institute for
Computational Sciences.

\appendix
\section{Fluid Limits of the Magnetic Variance Anisotropy}
Here we provide analytical derivations for the $\mathcal{A}_m$ from fluid theories. Without loss of generality, we will assume an equilibrium magnetic field $\bB_0 = B_0 \hat{\bz}$ and $\bk = (k_\perp, 0, k_\parallel)$ for all of the systems considered. For the purposes of constructing the $\mathcal{A}_m$, we need only consider the eigenfunctions of the magnetic field.

\subsection{Magnetohydrodynamics}\label{app:mhd}
 The Magnetohydrodynamic (MHD) equations can be written as
\begin{equation}
\frac{\partial \rho}{\partial t} + \nabla \cdot (\rho \bu) = 0,
\end{equation} 
\begin{equation}
\rho \frac{d \bu}{dt} = -\nabla p +\frac{1}{4\pi} (\nabla \times \bB ) \times \bB,
\end{equation} 
\begin{equation}
\frac{d}{dt} \left(\frac{p}{\rho^\gamma}\right) = 0,
\end{equation} 
\begin{equation}
c \bE + \bu \times \bB = 0,
\end{equation}
\begin{equation}\label{eq:divB}
\nabla \cdot \bB = 0,
\end{equation} 
\begin{equation}
\frac{\partial \bB}{\partial t} + c \nabla \times \bE = 0.
\end{equation}
The linear dispersion relation for this system is
\begin{equation}
\begin{split}
&(\omega^2 - v_A^2 k_\parallel^2)\left[\omega^4 - \omega^2 k^2(c_s^2 + v_A^2) + k_\parallel^2 k^2 v_A^2 c_s^2\right] =\\
&0,
\end{split}
\end{equation}
where the first term corresponds to the Alfv\'{e}n root and the two remaining roots correspond to the fast and slow compressible roots. Finally, the eigenfunctions for the system can be expressed as

\begin{equation}\label{eq:mhdBx}
\begin{split}
&(\omega^2 - v_A^2 k_\parallel^2) \frac{\delta B_x}{B_0} = \\
-k_\perp k_\parallel &\frac{\omega^2 (c_s^2 + v_A^2) -  c_s^2 v_A^2 k_\parallel^2}{\omega^2 - c_s^2 k_\parallel^2} \frac{\delta B_\parallel}{B_0}
\end{split}
\end{equation}
and
\begin{equation}
(\omega^2 - v_A^2 k_\parallel^2) \frac{\delta B_y}{B_0} = 0.
\end{equation}

The Alfv\'{e}n root corresponds to $\omega^2 = v_A^2 k_\parallel^2$, which implies $\delta B_\parallel = 0$. Therefore, the $\mathcal{A}_m$ for the MHD Alfv\'{e}n root is formally infinite.

The fast and slow compressible roots correspond to $\omega^2 \neq v_A^2 k_\parallel^2$, which implies $\delta B_y = 0$. Therefore, $\delta B_\perp^2 = \delta B_x^2$ and after manipulation equation~\eqref{eq:mhdBx} reduces to
\begin{equation}
\mathcal{A}_m = \frac{k_\parallel^2}{k_\perp^2}.
\end{equation}
Note, the $\mathcal{A}_m$ for the fast and slow roots can be trivially derived from equation~\eqref{eq:divB} since $\delta B_y = 0$.

\subsection{Electron MHD}\label{app:emhd}

Electron MHD (EMHD) is valid for $T_i \ll T_e$ ($\beta_i \gg 1$) and scales $k d_i > 1$ \citep{Kingsep:1990,Schekochihin:2009} and is most useful for describing the whistler mode. At these scales, the ions decouple from the magnetic field and are treated as being stationary within the framework of EMHD. Therefore, Ohm's law reduces to $\bE = -\bu_e \times B / c$. Inserting this form of Ohm's law into Faraday's law together with $\bJ = - n e \bu_e$, we obtain the EMHD equation
\begin{equation}
\frac{\partial \bB}{\partial t} + \nabla \times \left[ \left(\nabla \times \bB \right) \times \frac{c \bB}{4 \pi n e} \right] = 0.
\end{equation}

The linear dispersion relation for this equation is
\begin{equation}
\omega = \pm v_A k_\parallel d_i k,
\end{equation}
and the eigenfunctions for this system are
\begin{equation}
\delta B_x = -\frac{k_\parallel}{k_\perp} \delta B_\parallel
\end{equation}
and
\begin{equation}
\delta B_y = \pm i \frac{k}{k_\perp} \delta B_\parallel.
\end{equation}
Thus, the $\mathcal{A}_m$ for the whistler root is
\begin{equation}
\mathcal{A}_m = \frac{k_\parallel^2 + k^2}{k_\perp^2} = 1 + 2 \frac{k_\parallel^2}{k_\perp^2}.
\end{equation}

\subsection{Electron Reduced MHD}\label{app:ermhd}
Electron Reduced MHD (ERMHD) was introduced in \citet{Schekochihin:2009} and is the $k_\perp \rho_i \gg 1$, $m_e/m_i \ll 1$ limit of gyrokinetics. The system generalizes the EMHD equations for low-frequency, anisotropic ($k_\parallel \ll k_\perp$) fluctuations without assuming incompressibility, and kinetic Alfv\'{e}n waves are described well by ERMHD. The equations of ERMHD are most simply expressed in terms of scalar flux functions
\begin{equation}
v_A \frac{\delta \bB_\perp}{B_0} = \hat{\bz} \times \nabla_\perp \Psi
\end{equation}
and
\begin{equation}
\delta \bu_\perp = \hat{\bz} \times \nabla_\perp \Phi.
\end{equation}
The ERMHD equations are
\begin{equation}
\frac{\partial \Psi}{\partial t} = v_A \left(1 + \frac{T_e}{T_i}\right) \hat{\bz}\cdot\nabla\Phi,
\end{equation}
\begin{equation}
\frac{\partial \Phi}{\partial t} = -\frac{v_A}{ 2 + \beta_i \left(1 + \frac{T_e}{T_i}\right)} \hat{\bz}\cdot\nabla \left(\rho_i^2 \nabla_\perp^2 \Psi\right),
\end{equation}
\begin{equation}\label{eq:ermhdBz}
\frac{\delta B_\parallel}{B_0} = \sqrt{\beta_i} \left(1 + \frac{T_e}{T_i} \right) \frac{\Phi}{\rho_i v_A},
\end{equation}
\begin{equation}
\frac{\delta n_e}{n_0} = - \frac{2}{\sqrt{\beta_i}} \frac{\Phi}{\rho_i v_A},
\end{equation}
\begin{equation}
u_{\parallel e} = - \frac{\rho_i \nabla_\perp^2 \Psi}{\sqrt{\beta_i}},
\end{equation}
where $\hat{\bz} \cdot \nabla *= \partial */ \partial z + (1/v_A) \hat{\bz} \cdot \left(\nabla_\perp \Psi \times \nabla_\perp *\right)$. Note, the anisotropy of the system together with $\nabla \cdot \bB = 0$ implies $\delta B_x \ll \delta B_z \sim \delta B_y$, so we need only determine the eigenfunction for $\delta B_y$.

The linear dispersion relation for this system is
\begin{equation}
\omega = \pm \sqrt{\frac{1+T_e/T_i}{2 + \beta_i (1 + T_e/T_i)}} k_\perp \rho_i k_\parallel v_A
\end{equation}
and the eigenfunction is
\begin{equation}\label{eq:ermhdBy}
\begin{split}
&k_\perp \rho_i \Psi = \rho_i v_A \frac{\delta B_y}{B_0} = \\
\pm &\sqrt{\left(1 + \frac{T_e}{T_i}\right)\left[2 + \beta_i \left(1 + \frac{T_e}{T_i} \right) \right]} \Phi.
\end{split}
\end{equation}
Combining equations ~\eqref{eq:ermhdBz} and ~\eqref{eq:ermhdBy} yields the $\mathcal{A}_m$ for the kinetic Alfv\'{e}n wave.
\begin{equation}
\mathcal{A}_m = \frac{2 + \beta_i \left(1+T_e/T_i\right)}{\beta_i \left(1+T_e/T_i\right)}.
\end{equation}


\begin{thebibliography}{82}
\expandafter\ifx\csname natexlab\endcsname\relax\def\natexlab#1{#1}\fi

\bibitem[{{Alexandrova} {et~al.}(2011){Alexandrova}, {Lacombe}, {Mangeney}, \&
  {Grappin}}]{Alexandrova:2011}
{Alexandrova}, O., {Lacombe}, C., {Mangeney}, A., \& {Grappin}, R. 2011, ArXiv
  e-prints

\bibitem[{{Bale} {et~al.}(2005){Bale}, {Kellogg}, {Mozer}, {Horbury}, \&
  {Reme}}]{Bale:2005}
{Bale}, S.~D., {Kellogg}, P.~J., {Mozer}, F.~S., {Horbury}, T.~S., \& {Reme},
  H. 2005, Phys.~Rev.~Lett., 94, 215002

\bibitem[{{Barnes}(1966)}]{Barnes:1966}
{Barnes}, A. 1966, Phys.~Fluids, 9, 1483

\bibitem[{{Belcher} \& {Davis}(1971)}]{Belcher:1971}
{Belcher}, J.~W., \& {Davis}, L. 1971, J.~Geophys.~Res., 76, 3534

\bibitem[{{Boldyrev}(2005)}]{Boldyrev:2005}
{Boldyrev}, S. 2005, Astrophys.~J.~Lett., 626, L37

\bibitem[{{Boldyrev}(2006)}]{Boldyrev:2006}
---. 2006, Phys.~Rev.~Lett., 96, 115002

\bibitem[{{Boldyrev} {et~al.}(2011){Boldyrev}, {Perez}, {Borovsky}, \&
  {Podesta}}]{Boldyrev:2011}
{Boldyrev}, S., {Perez}, J.~C., {Borovsky}, J.~E., \& {Podesta}, J.~J. 2011,
  Astrophys.~J.~Lett., 741, L19

\bibitem[{{Bruno} \& {Carbone}(2005)}]{Bruno:2005}
{Bruno}, R., \& {Carbone}, V. 2005, Living Reviews in Solar Physics, 2

\bibitem[{{Burlaga} {et~al.}(1981){Burlaga}, {Sittler}, {Mariani}, \&
  {Schwenn}}]{Burlaga:1981}
{Burlaga}, L., {Sittler}, E., {Mariani}, F., \& {Schwenn}, R. 1981,
  J.~Geophys.~Res., 86, 6673

\bibitem[{{Burlaga} \& {Ogilvie}(1970)}]{Burlaga:1970}
{Burlaga}, L.~F., \& {Ogilvie}, K.~W. 1970, Solar Physics, 15, 61

\bibitem[{{Burlaga} \& {Ogilvie}(1973)}]{Burlaga:1973}
---. 1973, J.~Geophys.~Res., 78, 2028

\bibitem[{{Chandran} {et~al.}(2009){Chandran}, {Quataert}, {Howes}, {Xia}, \&
  {Pongkitiwanichakul}}]{Chandran:2009b}
{Chandran}, B.~D.~G., {Quataert}, E., {Howes}, G.~G., {Xia}, Q., \&
  {Pongkitiwanichakul}, P. 2009, Astrophys.~J., 707, 1668

\bibitem[{{Chen} {et~al.}(2011{\natexlab{a}}){Chen}, {Mallet}, {Schekochihin},
  {Horbury}, {Wicks}, \& {Bale}}]{Chen:2011}
{Chen}, C.~H.~K., {Mallet}, A., {Schekochihin}, A.~A., {et~al.}
  2011{\natexlab{a}}, ArXiv e-prints

\bibitem[{{Chen} {et~al.}(2011{\natexlab{b}}){Chen}, {Mallet}, {Yousef},
  {Schekochihin}, \& {Horbury}}]{Chen:2011a}
{Chen}, C.~H.~K., {Mallet}, A., {Yousef}, T.~A., {Schekochihin}, A.~A., \&
  {Horbury}, T.~S. 2011{\natexlab{b}}, Mon.~Not.~Roy.~Astron.~Soc., 415, 3219

\bibitem[{{Cho} \& {Lazarian}(2002)}]{Cho:2002a}
{Cho}, J., \& {Lazarian}, A. 2002, Phys.~Rev.~Lett., 88, 245001

\bibitem[{{Cho} \& {Lazarian}(2003)}]{Cho:2003}
---. 2003, Mon.~Not.~Roy.~Astron.~Soc., 345, 325

\bibitem[{{Cho} \& {Lazarian}(2004)}]{Cho:2004}
---. 2004, Astrophys.~J.~Lett., 615, L41

\bibitem[{Cho \& Vishniac(2000)}]{Cho:2000}
Cho, J., \& Vishniac, E.~T. 2000, Astrophys.~J., 539, 273

\bibitem[{{Dastgeer} {et~al.}(2000){Dastgeer}, {Das}, {Kaw}, \&
  {Diamond}}]{Dastgeer:2000}
{Dastgeer}, S., {Das}, A., {Kaw}, P., \& {Diamond}, P.~H. 2000, Phys.~Plasmas,
  7, 571

\bibitem[{{Fedorov} {et~al.}(2011){Fedorov}, {Opitz}, {Sauvaud}, {Luhmann},
  {Curtis}, \& {Larson}}]{Fedorov:2011}
{Fedorov}, A., {Opitz}, A., {Sauvaud}, J.-A., {et~al.} 2011, Space Sci.~Rev.,
  161, 49

\bibitem[{{Galtier}(2006)}]{Galtier:2006}
{Galtier}, S. 2006, J.~Plasma Phys., 72, 721

\bibitem[{{Galtier} \& {Bhattacharjee}(2003)}]{Galtier:2003}
{Galtier}, S., \& {Bhattacharjee}, A. 2003, Phys.~Plasmas, 10, 3065

\bibitem[{{Galvin} {et~al.}(2008){Galvin}, {Kistler}, {Popecki}, {Farrugia},
  {Simunac}, {Ellis}, {M{\"o}bius}, {Lee}, {Boehm}, {Carroll}, {Crawshaw},
  {Conti}, {Demaine}, {Ellis}, {Gaidos}, {Googins}, {Granoff}, {Gustafson},
  {Heirtzler}, {King}, {Knauss}, {Levasseur}, {Longworth}, {Singer}, {Turco},
  {Vachon}, {Vosbury}, {Widholm}, {Blush}, {Karrer}, {Bochsler}, {Daoudi},
  {Etter}, {Fischer}, {Jost}, {Opitz}, {Sigrist}, {Wurz}, {Klecker}, {Ertl},
  {Seidenschwang}, {Wimmer-Schweingruber}, {Koeten}, {Thompson}, \&
  {Steinfeld}}]{Galvin:2008}
{Galvin}, A.~B., {Kistler}, L.~M., {Popecki}, M.~A., {et~al.} 2008, Space
  Sci.~Rev., 136, 437

\bibitem[{{Gary} {et~al.}(1998){Gary}, {Li}, {O'Rourke}, \&
  {Winske}}]{Gary:1998}
{Gary}, S.~P., {Li}, H., {O'Rourke}, S., \& {Winske}, D. 1998,
  J.~Geophys.~Res., 103, 14567

\bibitem[{{Gary} \& {Smith}(2009)}]{Gary:2009}
{Gary}, S.~P., \& {Smith}, C.~W. 2009, J.~Geophys.~Res., 114, 12105

\bibitem[{Goldreich \& Sridhar(1995)}]{Goldreich:1995}
Goldreich, P., \& Sridhar, S. 1995, Astrophys.~J., 438, 763

\bibitem[{Goldreich \& Sridhar(1997)}]{Goldreich:1997}
---. 1997, Astrophys.~J., 485, 680

\bibitem[{{Grappin} {et~al.}(1990){Grappin}, {Mangeney}, \&
  {Marsch}}]{Grappin:1990}
{Grappin}, R., {Mangeney}, A., \& {Marsch}, E. 1990, J.~Geophys.~Res., 95, 8197

\bibitem[{{Hamilton} {et~al.}(2008){Hamilton}, {Smith}, {Vasquez}, \&
  {Leamon}}]{Hamilton:2008}
{Hamilton}, K., {Smith}, C.~W., {Vasquez}, B.~J., \& {Leamon}, R.~J. 2008,
  J.~Geophys.~Res., 113, A01106

\bibitem[{{He} {et~al.}(2012){He}, {Tu}, {Marsch}, \& {Yao}}]{He:2012}
{He}, J., {Tu}, C., {Marsch}, E., \& {Yao}, S. 2012, Astrophys.~J.~Lett., 745,
  L8

\bibitem[{{Hirose} {et~al.}(2004){Hirose}, {Ito}, {Mahajan}, \&
  {Ohsaki}}]{Hirose:2004}
{Hirose}, A., {Ito}, A., {Mahajan}, S.~M., \& {Ohsaki}, S. 2004, Physics
  Letters A, 330, 474

\bibitem[{{Horbury} {et~al.}(2008){Horbury}, Forman, \& Oughton}]{Horbury:2008}
{Horbury}, T.~S., Forman, M., \& Oughton, S. 2008, Phys.~Rev.~Lett., 101,
  175005

\bibitem[{Howes(2009)}]{Howes:2009b}
Howes, G.~G. 2009, Nonlin.~Proc.~Geophys., 16, 219

\bibitem[{{Howes} {et~al.}(2011{\natexlab{a}}){Howes}, {Bale}, {Klein}, {Chen},
  {Salem}, \& {TenBarge}}]{Howes:2011a}
{Howes}, G.~G., {Bale}, S.~D., {Klein}, K.~G., {et~al.} 2011{\natexlab{a}},
  ArXiv e-prints, submitted to Astrophys. J. Lett.

\bibitem[{{Howes} {et~al.}(2006){Howes}, {Cowley}, {Dorland}, {Hammett},
  {Quataert}, \& {Schekochihin}}]{Howes:2006}
{Howes}, G.~G., {Cowley}, S.~C., {Dorland}, W., {et~al.} 2006, Astrophys.~J.,
  651, 590

\bibitem[{{Howes} {et~al.}(2008){Howes}, {Cowley}, {Dorland}, {Hammett},
  {Quataert}, \& {Schekochihin}}]{Howes:2008b}
---. 2008, J.~Geophys.~Res., 113, A05103

\bibitem[{{Howes} \& {Quataert}(2010)}]{Howes:2010a}
{Howes}, G.~G., \& {Quataert}, E. 2010, Astrophys.~J.~Lett., 709, L49

\bibitem[{{Howes} {et~al.}(2011{\natexlab{b}}){Howes}, {Tenbarge}, \&
  {Dorland}}]{Howes:2011c}
{Howes}, G.~G., {Tenbarge}, J.~M., \& {Dorland}, W. 2011{\natexlab{b}},
  Phys.~Plasmas, 18, 102305

\bibitem[{{Howes} {et~al.}(2011{\natexlab{c}}){Howes}, {Tenbarge}, {Dorland},
  {Quataert}, {Schekochihin}, {Numata}, \& {Tatsuno}}]{Howes:2011b}
{Howes}, G.~G., {Tenbarge}, J.~M., {Dorland}, W., {et~al.} 2011{\natexlab{c}},
  Phys.~Rev.~Lett., 107, 035004

\bibitem[{{Howes} {et~al.}(2012){Howes}, {TenBarge}, \& {Klein}}]{Howes:2012a}
{Howes}, G.~G., {TenBarge}, J.~M., \& {Klein}, K.~G. 2012, Astrophys.~J., in
  preparation

\bibitem[{{Ito} {et~al.}(2004){Ito}, {Hirose}, {Mahajan}, \&
  {Ohsaki}}]{Ito:2004}
{Ito}, A., {Hirose}, A., {Mahajan}, S.~M., \& {Ohsaki}, S. 2004, Phys.~Plasmas,
  11, 5643

\bibitem[{{Kellogg} \& {Horbury}(2005)}]{Kellogg:2005}
{Kellogg}, P.~J., \& {Horbury}, T.~S. 2005, Ann.~Geophys., 23, 3765

\bibitem[{{Kingsep} {et~al.}(1990){Kingsep}, {Chukbar}, \&
  {Yankov}}]{Kingsep:1990}
{Kingsep}, A.~S., {Chukbar}, K.~V., \& {Yankov}, V.~V. 1990, Rev. Plasma Phys.,
  16, 243

\bibitem[{{Klein} {et~al.}(2012{\natexlab{a}}){Klein}, {Howes}, \&
  {TenBarge}}]{Klein:2012a}
{Klein}, K.~G., {Howes}, G.~G., \& {TenBarge}, J.~M. 2012{\natexlab{a}},
  Phys.~Plasmas, in preparation

\bibitem[{{Klein} {et~al.}(2012{\natexlab{b}}){Klein}, {Howes}, {TenBarge},
  {Bale}, {Chen}, \& {Salem}}]{Klein:2012}
{Klein}, K.~G., {Howes}, G.~G., {TenBarge}, J.~M., {et~al.} 2012{\natexlab{b}},
  Astrophys.~J., submitted

\bibitem[{Leamon {et~al.}(1998)Leamon, Smith, Ness, Matthaeus, \&
  Wong}]{Leamon:1998a}
Leamon, R.~J., Smith, C.~W., Ness, N.~F., Matthaeus, W.~H., \& Wong, H.~K.
  1998, J.~Geophys.~Res., 103, 4775

\bibitem[{{Li} \& {Habbal}(2001)}]{LiHabbal:2001}
{Li}, X., \& {Habbal}, S.~R. 2001, J.~Geophys.~Res., 106, 10669

\bibitem[{Lithwick \& Goldreich(2001)}]{Lithwick:2001}
Lithwick, Y., \& Goldreich, P. 2001, Astrophys.~J., 562, 279

\bibitem[{{Luo} \& {Wu}(2010)}]{Luo:2010}
{Luo}, Q.~Y., \& {Wu}, D.~J. 2010, Astrophys.~J.~Lett., 714, L138

\bibitem[{Maron \& Goldreich(2001)}]{Maron:2001}
Maron, J., \& Goldreich, P. 2001, Astrophys.~J., 554, 1175

\bibitem[{{Mason} {et~al.}(2006){Mason}, {Cattaneo}, \&
  {Boldyrev}}]{Mason:2006}
{Mason}, J., {Cattaneo}, F., \& {Boldyrev}, S. 2006, Phys.~Rev.~Lett., 97,
  255002

\bibitem[{{Mason} {et~al.}(2008){Mason}, {Cattaneo}, \&
  {Boldyrev}}]{Mason:2008}
---. 2008, Phys.~Rev.~E, 77, 036403

\bibitem[{{Matthaeus} {et~al.}(1995){Matthaeus}, {Bieber}, \&
  {Zank}}]{Matthaeus:1995}
{Matthaeus}, W.~H., {Bieber}, J.~W., \& {Zank}, G.~P. 1995, Rev. Geophys., 33,
  609

\bibitem[{{Narita} \& {Gary}(2010)}]{Narita:2010a}
{Narita}, Y., \& {Gary}, S.~P. 2010, Annales Geophysicae, 28, 597

\bibitem[{{Newbury} {et~al.}(1998){Newbury}, {Russell}, {Phillips}, \&
  {Gary}}]{Newbury:1998}
{Newbury}, J.~A., {Russell}, C.~T., {Phillips}, J.~L., \& {Gary}, S.~P. 1998,
  J.~Geophys.~Res., 103, 9553

\bibitem[{{Numata} {et~al.}(2010){Numata}, {Howes}, {Tatsuno}, {Barnes}, \&
  {Dorland}}]{Numata:2010}
{Numata}, R., {Howes}, G.~G., {Tatsuno}, T., {Barnes}, M., \& {Dorland}, W.
  2010, J.~Comp.~Phys., 229, 9347

\bibitem[{{Osherovich} {et~al.}(1993){Osherovich}, {Farrugia}, {Burlaga},
  {Lepping}, {Fainberg}, \& {Stone}}]{Osherovich:1993}
{Osherovich}, V.~A., {Farrugia}, C.~J., {Burlaga}, L.~F., {et~al.} 1993, 98,
  15331

\bibitem[{{Podesta}(2009)}]{Podesta:2009a}
{Podesta}, J.~J. 2009, Astrophys.~J., 698, 986

\bibitem[{{Podesta} \& {Borovsky}(2010)}]{Podesta:2010b}
{Podesta}, J.~J., \& {Borovsky}, J.~E. 2010, Phys.~Plasmas, 17, 112905

\bibitem[{{Podesta} \& {Gary}(2011)}]{Podesta:2011a}
{Podesta}, J.~J., \& {Gary}, S.~P. 2011, Astrophys.~J., 734, 15

\bibitem[{{Podesta} \& {TenBarge}(2012)}]{Podesta:2012}
{Podesta}, J.~J., \& {TenBarge}, J.~M. 2012, J.~Geophys.~Res., submitted

\bibitem[{{Quataert}(1998)}]{Quataert:1998}
{Quataert}, E. 1998, Astrophys.~J., 500, 978

\bibitem[{{Richardson} {et~al.}(1997){Richardson}, {Farrugia}, \&
  {Cane}}]{Richardson:1997}
{Richardson}, I.~G., {Farrugia}, C.~J., \& {Cane}, H.~V. 1997,
  J.~Geophys.~Res., 102, 4691

\bibitem[{{Roberts}(1990)}]{Roberts:1990}
{Roberts}, D.~A. 1990, J.~Geophys.~Res., 95, 1087

\bibitem[{{Sahraoui} {et~al.}(2011){Sahraoui}, {Belmont}, \&
  {Goldstein}}]{Sahraoui:2011}
{Sahraoui}, F., {Belmont}, G., \& {Goldstein}, M. 2011, ArXiv e-prints

\bibitem[{{Sahraoui} {et~al.}(2010){Sahraoui}, {Goldstein}, {Belmont}, {Canu},
  \& {Rezeau}}]{Sahraoui:2010b}
{Sahraoui}, F., {Goldstein}, M.~L., {Belmont}, G., {Canu}, P., \& {Rezeau}, L.
  2010, Phys.~Rev.~Lett., 105, 131101

\bibitem[{{Saito} {et~al.}(2010){Saito}, {Gary}, \& {Narita}}]{Saito:2010}
{Saito}, S., {Gary}, S.~P., \& {Narita}, Y. 2010, Phys.~Plasmas, 17, 122316

\bibitem[{{Salem} {et~al.}(2012){Salem}, {Howes}, {Sundkvist}, {Bale},
  {Chaston}, {Chen}, \& {Mozer}}]{Salem:2012}
{Salem}, C.~S., {Howes}, G.~G., {Sundkvist}, D., {et~al.} 2012,
  Astrophys.~J.~Lett., 745, L9

\bibitem[{{Schekochihin} {et~al.}(2009){Schekochihin}, {Cowley}, {Dorland},
  {Hammett}, {Howes}, {Quataert}, \& {Tatsuno}}]{Schekochihin:2009}
{Schekochihin}, A.~A., {Cowley}, S.~C., {Dorland}, W., {et~al.} 2009,
  Astrophys.~J.~Supp., 182, 310

\bibitem[{{Smith} {et~al.}(1998){Smith}, {L'Heureux}, {Ness}, {Acu{\~n}a},
  {Burlaga}, \& {Scheifele}}]{Smith:1998}
{Smith}, C.~W., {L'Heureux}, J., {Ness}, N.~F., {et~al.} 1998, Space Sci.~Rev.,
  86, 613

\bibitem[{{Smith} {et~al.}(2006){Smith}, {Vasquez}, \&
  {Hamilton}}]{Smith:2006a}
{Smith}, C.~W., {Vasquez}, B.~J., \& {Hamilton}, K. 2006, J.~Geophys.~Res.,
  111, A09111

\bibitem[{{Smith} {et~al.}(2012){Smith}, {Vasquez}, \& {Hollweg}}]{Smith:2012}
{Smith}, C.~W., {Vasquez}, B.~J., \& {Hollweg}, J.~V. 2012, Astrophys.~J., 745,
  8

\bibitem[{{Stix}(1992)}]{Stix:1992}
{Stix}, T.~H. 1992, {Waves in Plasmas} (New York: American Institute of
  Physics)

\bibitem[{{Svidzinski} {et~al.}(2009){Svidzinski}, {Li}, {Rose}, {Albright}, \&
  {Bowers}}]{Svidzinski:2009}
{Svidzinski}, V.~A., {Li}, H., {Rose}, H.~A., {Albright}, B.~J., \& {Bowers},
  K.~J. 2009, Phys.~Plasmas, 16, 122310

\bibitem[{{TenBarge} \& {Howes}(2012{\natexlab{a}})}]{TenBarge:2012c}
{TenBarge}, J.~M., \& {Howes}, G.~G. 2012{\natexlab{a}}, Phys.~Rev.~Lett., submitted

\bibitem[{{TenBarge} \& {Howes}(2012{\natexlab{b}})}]{TenBarge:2011a}
---. 2012{\natexlab{b}}, Phys.~Plasmas, 19, 055901

\bibitem[{{TenBarge} {et~al.}(2012){TenBarge}, {Howes}, {Dorland}, \&
  {Hammett}}]{TenBarge:2012b}
{TenBarge}, J.~M., {Howes}, G.~G., {Dorland}, W., \& {Hammett}, G.~W. 2012,
  J.~Comp.~Phys., in preparation

\bibitem[{{Tu} \& {Marsch}(1995)}]{Tu:1995}
{Tu}, C.-Y., \& {Marsch}, E. 1995, Space Science Reviews, 73, 1

\bibitem[{{Turner} {et~al.}(2011){Turner}, {Gogoberidze}, {Chapman}, {Hnat}, \&
  {M{\"u}ller}}]{Turner:2011}
{Turner}, A.~J., {Gogoberidze}, G., {Chapman}, S.~C., {Hnat}, B., \&
  {M{\"u}ller}, W.-C. 2011, Phys.~Rev.~Lett., 107, 095002

\bibitem[{{Vellante} \& {Lazarus}(1987)}]{Vellante:1987}
{Vellante}, M., \& {Lazarus}, A.~J. 1987, J.~Geophys.~Res., 92, 9893

\bibitem[{{Wicks} {et~al.}(2012){Wicks}, {Forman}, {Horbury}, \&
  {Oughton}}]{Wicks:2012}
{Wicks}, R.~T., {Forman}, M.~A., {Horbury}, T.~S., \& {Oughton}, S. 2012,
  Astrophys.~J., 746, 103

\bibitem[{{Wicks} {et~al.}(2010){Wicks}, {Horbury}, {Chen}, \&
  {Schekochihin}}]{Wicks:2010a}
{Wicks}, R.~T., {Horbury}, T.~S., {Chen}, C.~H.~K., \& {Schekochihin}, A.~A.
  2010, Mon.~Not.~Roy.~Astron.~Soc., 407, L31

\end{thebibliography}

\end{document}